\documentclass[12pt]{article}
\usepackage{amsmath,amssymb}
\usepackage{bm}
\usepackage{color}
\usepackage{graphicx}
\usepackage{cite}
\usepackage{mathtools}

\usepackage{slashed}
\usepackage{here}

\numberwithin{equation}{section}

\begin{document}

\title{
\begin{flushright}
\ \\*[-80pt]
\begin{minipage}{0.23\linewidth}
\normalsize
EPHOU-26-04\\
KYUSHU-HET-358\\*[50pt]
\end{minipage}
\end{flushright}
{\Large \bf
Massive modes on magnetized blow-up manifold of $T^2/\mathbb{Z}_N$
\\*[20pt]}}

\author{
Tatsuo Kobayashi $^{1}$,
~Hajime Otsuka $^{2,3}$, 
and
~Hikaru Uchida $^{4}$
\\*[20pt]
\centerline{
\begin{minipage}{\linewidth}
\begin{center}
$^1${\it \normalsize
Department of Physics, Hokkaido University, Sapporo 060-0810, Japan} \\*[5pt]
$^2${\it \normalsize
Department of Physics, Kyushu University, 744 Motooka, Nishi-ku, Fukuoka 819-0395, Japan} \\*[5pt]
$^3${\it \normalsize
Quantum and Spacetime Research Institute (QuaSR), Kyushu University, 744 Motooka, Nishi-ku, Fukuoka, 819-0395, Japan} \\*[5pt]
$^4${\it \normalsize
National Institute of Technology, Hakodate College, Hakodate 042-0953, Japan} \\*[5pt]
\end{center}
\end{minipage}}
\\*[50pt]}

\date{
\centerline{\small \bf Abstract}
\begin{minipage}{0.9\linewidth}
\medskip
\medskip
\small
We study massive modes on a magnetized blow-up manifold of $T^2/\mathbb{Z}_N$.
The blow-up manifold can be constructed by appropriately replacing orbifold singular points with a part of $S^2$. 
To ensure a smooth connection between the massive modes on magnetized $T^2/\mathbb{Z}_N$ orbifold and those on magnetized $S^2$, it is required that not only the total magnetic flux as well as the total curvature but also the effective magnetic flux on the connected line remain invariant under the blow-up procedure. 
Furthermore, we find that the number of the localized modes at each orbifold singular point increases by one for each unit increment of the mass level.
\end{minipage}
}

\begin{titlepage}
\maketitle
\thispagestyle{empty}
\end{titlepage}

\newpage



\section{Introduction}
\label{sec:introduction}

Calabi-Yau manifold compactification is one of the most attractive compactifications to obtain a four-dimensional chiral theory including the standard model in superstring theory \cite{Candelas:1985en}.
However, it is difficult to calculate analytically all couplings in the four-dimensional low effective field theory derived from Calabi-Yau manifolds in general.

Toroidal orbifold compactifications~\cite{Dixon:1985jw,Dixon:1986jc}, which can be viewed as singular limits of certain Calabi-Yau manifolds, are of particular interest because their four-dimensional effective field theory can be calculated analytically in principle. 
Furthermore, torus and orbifold models with magnetic flux backgrounds are interesting. 
Multi-generational chiral fermions appear~\cite{Bachas:1995ik,Blumenhagen:2000wh,Angelantonj:2000hi,Blumenhagen:2000ea,Cremades:2004wa,Abe:2008fi,Abe:2013bca,Abe:2014noa,Kobayashi:2017dyu,Sakamoto:2020pev,Kikuchi:2022lfv,Kikuchi:2022psj,Kikuchi:2023awm}\footnote{Three generational models have been studied in Refs.~\cite{Abe:2008sx,Abe:2015yva,Hoshiya:2020hki}.} and their couplings can be calculated through the overlap integration of their wave functions~\cite{Cremades:2004wa,Fujimoto:2016zjs,Abe:2009dr,Kobayashi:2015siy,Hoshiya:2021nux,Jeric:2025exr}.
Indeed, realistic quark and lepton masses, their flavor mixing angles, and the CP phase have been obtained in Refs.~\cite{Fujimoto:2016zjs,Abe:2012fj,Abe:2014vza,Kobayashi:2016qag,Buchmuller:2017vho,Buchmuller:2017vut,Kikuchi:2021yog,Kikuchi:2022geu,Hoshiya:2022qvr}.
Recently, it has been found that their flavor structure can be related to the modular symmetry~\cite{Kobayashi:2017dyu,Hoshiya:2020hki,Hoshiya:2021nux,Kobayashi:2018rad,Kikuchi:2020frp,Kikuchi:2021ogn,Kobayashi:2018bff,Kikuchi:2022bkn,Kikuchi:2023clx,Kikuchi:2023awe,Kobayashi:2024ysa,Kobayashi:2024hkk,Jeric:2025iwk,Ohki:2020bpo} and non-invertible symmetry~\cite{Kobayashi:2024yqq}.
In addition to zero modes, massive modes have also been studied~\cite{Berasaluce-Gonzalez:2012abm,Hamada:2012wj,Abe:2013bca}.

One can blow-up orbifold singular points to construct appropriately smooth manifolds such as Calabi-Yau manifolds, where topological aspects are the same as in the orbifold limit. 
By use of such blow-up procedures, we can analytically discuss the four-dimensional low energy effective field theory near the orbifold limit in the whole moduli space of Calabi-Yau manifolds. 
The blow-up manifold of $\mathbb{C}^N/\mathbb{Z}_N$ with $N \geq 2$ has been studied in Refs.~\cite{GrootNibbelink:2007lua,Leung:2019oln}, where it is constructed by replacing the singular points with the Eguchi-Hanson spaces~\cite{Eguchi:1978xp}. 
However, wave functions and their couplings on these blow-up manifolds have not been found yet.

On the other hand, the blow-up manifold of magnetized $T^2/\mathbb{Z}_N$ orbifold has been studied in Refs.~\cite{Kobayashi:2019fma,Kobayashi:2019gyl,Kobayashi:2022tti}, where it is constructed by replacing the singular points with a part of $S^2$ appropriately. 
Wave functions themselves on the whole $S^2$ were studied with a uniform magnetic flux background in Ref.~\cite{Conlon:2008qi} and a vortex background in Ref.~\cite{Dolan:2020sjq}. 
(See also Ref.~\cite{Dolan:2003bj}.)
In Refs.~\cite{Kobayashi:2019fma,Kobayashi:2019gyl,Kobayashi:2022tti}, 
we smoothly connect the zero-mode wave functions on the magnetized $T^2/\mathbb{Z}_N$ orbifold outside of the singular points and 
a part of $S^2$ with the uniform magnetic flux in order to construct zero-mode wave functions on the blow-up manifold. 
We also studied their couplings analytically and numerically.
Yukawa couplings depend on the blow-up radius, and that would be important to realize quark and lepton masses and their mixing angles. 
In particular, Ref.~\cite{Kobayashi:2022tti} showed that the magnetic fluxes inserted on orbifold singular points, called localized fluxes, generate additional zero-modes localized at the singular points, called localized modes, through the index theorem on the blow-up manifold. 
It is important to extend these analyses to massive modes, which may affect the four-dimensional low energy effective field theory, e.g. through loop effects such as threshold corrections to gauge and Yukawa couplings \cite{Dixon:1990pc,Antoniadis:1992pm}.

In this paper, in order to construct the whole system of the magnetized blow-up manifold compactification, we study not only zero modes but also massive modes on the magnetized blow-up manifold of $T^2/\mathbb{Z}_N$ orbifold by connecting massive modes on magnetized $T^2/\mathbb{Z}_N$ orbifold~\cite{Abe:2013bca} after the singular gauge transformation~\cite{Lee:2003mc,Buchmuller:2015eya,Buchmuller:2018lkz,Polchinski:1998rq,Kobayashi:2022tti} to massive modes on magnetized $S^2$ with a vortex~\cite{Dolan:2020sjq} smoothly. 
We note that a suitable vortex has to be introduced to connect massive modes smoothly, although we do not need it to connect only the zero-modes. 
By comparing the degenerate number of massive modes on magnetized $S^2$~\cite{Dolan:2020sjq} with that of massive modes on magnetized $T^2/\mathbb{Z}_N$ orbifold~\cite{Abe:2013bca}, which correspond to bulk modes, we find that additional massive modes appear locally at the orbifold singular points, which correspond to localized modes. 
This behavior is analogous to heterotic orbifold models, where there are towers of massive modes on orbifold fixed points \cite{Choi:2020dws}, as well as intersecting D-brane models, where towers of massive modes exist at the intersecting points \cite{Aldazabal:2000cn}.

This paper is organized as follows.
In section~\ref{sec:T2ZN}, we discuss massive modes on the magnetized $T^2/\mathbb{Z}_N$ orbifold.
First, we review massive modes on a magnetized $T^2$ in subsection~\ref{subsec:T2}. 
Then, we review massive modes on magnetized $T^2/\mathbb{Z}_N$ orbifold in 
subsection~\ref{subsec:T2ZN}. 
In subsection~\ref{subsec:singulargauge}, we study the massive modes after the singular gauge transformation.
We note that the gamma matrix factor is also affected by the singular gauge transformation. 
Section~\ref{sec:S2} is devoted to massive modes on the magnetized $S^2$. 
We discuss the cases without and with a vortex in subsections~\ref{subsec:withoutvortex} and~\ref{subsec:withvortex}, respectively. 
We comment on the angular momentum in subsection~\ref{subsec:AM}. 
In section~\ref{sec:blowupmanifold}, we study massive modes on the blow-up manifold of $T^2/\mathbb{Z}_N$ orbifold.
Based on the junction conditions, we find that not only the total magnetic flux as well as the total curvature but also the effective flux density on the connected line do not change under the blow-up procedure. 
Then, the massive modes on the magnetized $T^2/\mathbb{Z}_N$ orbifold can be smoothly connected to those on the magnetized $S^2$ by applying appropriate magnetic flux, vortex, and coefficients.
In section~\ref{sec:localizedmode}, we study the massive modes associated with localized modes. 
We conclude this paper in section~\ref{sec:conclusion}.
In Appendix~\ref{app:AM}, we define the angular momentum operators on the magnetized $T^2/\mathbb{Z}_N$ orbifold derived from those on the magnetized $S^2$ through the blow-up procedure. 
Explicit forms of the wave functions on magnetized $T^2/\mathbb{Z}_N$ orbifold and magnetized $S^2$ are provided in Appendices~\ref{app:T2ZN} and \ref{app:S2}, respectively.


\section{Magnetized $T^2/\mathbb{Z}_N$ orbifold}
\label{sec:T2ZN}

In this section, we review on magnetized $T^2/\mathbb{Z}_N$ models.
In subsection~\ref{subsec:singulargauge}, in particular, we modify the construction of excited states on magnetized $T^2/\mathbb{Z}_N$ and their mass eigenvalues by introducing the singular gauge transformation.


\subsection{Magnetized $T^2$}
\label{subsec:T2}

First, we review on magnetized $T^2$ models~\cite{Cremades:2004wa,Hamada:2012wj}.
A two-dimensional torus, $T^2$, can be constructed by dividing a complex plane, $\mathbb{C}$, by a two-dimensional lattice, $\Lambda$,~i.e. $T^2 \simeq \mathbb{C}/\Lambda$.
We write the coordinate of $T^2$ as $z$ and it satisfies $z+1 \sim z+\tau \sim z$, where $\tau$ denotes the complex structure modulus.
The metric of $T^2$ is written by
\begin{align}
    ds^2 = 2h_{z\bar{z}} dz d\bar{z} = dz d\bar{z},
    \label{eq:T2metric}
\end{align}
and the area of $T^2$ is ${\rm Im}\tau$.
The gamma matrices, $\gamma^{z}$ and $\gamma^{\bar{z}}$, are given by
\begin{align}
    \gamma^{z} =
    \begin{pmatrix}
        0 & 2 \\
        0 & 0
    \end{pmatrix},
    \quad
    \gamma^{\bar{z}} =
    \begin{pmatrix}
        0 & 0 \\
        2 & 0
    \end{pmatrix},
    \label{eq:gamma}
\end{align}
which satisfy $\{ \gamma^{z}, \gamma^{\bar{z}} \} = 2h^{z\bar{z}}$.
Here, $h^{z\bar{z}}$ denotes the inverse of $h_{z\bar{z}}$.
On the $T^2$, the $U(1)$ magnetic flux,
\begin{align}
    \frac{F}{2\pi} = \frac{i}{2} \frac{M}{{\rm Im}\tau} dz \land d\bar{z},
    \label{eq:F}
\end{align}
is inserted, where it satisfies $(2\pi)^{-1} \int_{T^2} F = M \in \mathbb{Z}$.
Hereafter, we consider $M>0$.
The magnetic flux is induced by the gauge potential,
\begin{align}
    A = -\frac{i}{2} \frac{\pi M}{{\rm Im}\tau}\bar{z} dz + \frac{i}{2} \frac{\pi M}{{\rm Im}\tau}z d\bar{z}.
    \label{eq:A}
\end{align}
Note that we do not consider Wilson line phases since they can be converted into Scherk-Schwarz phases.
The covariant derivatives are written by $D_{z} = \partial_{z} - iqA_{z}$ and $D_{\bar{z}} = \partial_{\bar{z}} - iqA_{\bar{z}}$, where $q$ denotes the $U(1)$ charge of a field.
In the following, we consider a two-dimensional spinor, $\psi_{T^2}^{M}(z)={^t}(\psi_{T^2,+}^{M}(z), \psi_{T^2,-}^{M}(z))$, on the magnetized $T^2$ with $U(1)$ unit charge $q=1$.
The spinor satisfies the following Dirac equation:
\begin{align}
    \begin{array}{l}
        -i {\cal D}^{\dagger} \psi_{T^2,-,n}^{M}(z) = \psi_{T^2,+,n}^{M}(z), \\
        i {\cal D} \psi_{T^2,+,n}^{M}(z) = {\cal M}_n^2 \psi_{T^2,-,n}^{M}(z),
    \end{array}
    \label{eq:DiraceqT2}
\end{align}
where ${\cal D}$ and ${\cal D}^{\dagger}$ are defined from the Dirac operator,
\begin{align}
    i \slashed{D} \equiv
    \begin{pmatrix}
        0 & -i{\cal D}^{\dagger} \\
        i{\cal D} & 0
    \end{pmatrix}
     =
    \begin{pmatrix}
        0 & 2iD_{z} \\
        2iD_{\bar{z}} & 0
    \end{pmatrix}
    =
    \begin{pmatrix}
        0 & 2i(\partial_{z}-\frac{\pi M}{2{\rm Im}\tau}\bar{z}) \\
        2i(\partial_{\bar{z}}+\frac{\pi M}{2{\rm Im}\tau}z) & 0
    \end{pmatrix},
    \label{eq:DiracopT2}
\end{align}
and ${\cal M}_{n}$ denotes the mass eigenvalue of the spinor with level $n$.\footnote{In this paper, we define the spinor such that the mass eigenvalue only appears in the second equation in Eq.~(\ref{eq:DiraceqT2}).}
It also satisfies the following boundary conditions related to $T^2$ translations,
\begin{align}
    \begin{array}{ll}
        \psi_{T^2,\pm,n}^{(\alpha_1,\alpha_{\tau}),M}(z+1) = e^{2\pi i \alpha_1} e^{i \chi_1(z)} \psi_{T^2,\pm,n}^{(\alpha_1,\alpha_{\tau}),M}(z), & \chi_1(z) = \pi M\frac{{\rm Im}z}{{\rm Im}\tau}, \\
        \psi_{T^2,\pm,n}^{(\alpha_1,\alpha_{\tau}),M}(z+\tau) = e^{2\pi i \alpha_{\tau}} e^{i \chi_{\tau}(z)} \psi_{T^2,\pm,n}^{(\alpha_1,\alpha_{\tau}),M}(z), & \chi_{\tau}(z) = \pi M\frac{{\rm Im}(\bar{\tau}z)}{{\rm Im}\tau},
    \end{array}
    \label{eq:BCT2}
\end{align}
where $(\alpha_1, \alpha_{\tau})$ denote the Scherk-Schwarz phases.
Under the boundary conditions in Eq.~(\ref{eq:BCT2}), the solutions of the Dirac equation in Eq.~(\ref{eq:DiraceqT2}) are obtained as
\begin{align}
    &\ \psi_{T^2,+,n}^{(\alpha_1,\alpha_{\tau}),M,j}(z) \equiv \psi_{T^2,n}^{(\alpha_1,\alpha_{\tau}),M,j}(z) = (-i {\cal D}^{\dagger})^{n} \psi_{T^2,0}^{(\alpha_1,\alpha_{\tau}),M,j}(z),
    \label{eq:psiT2+n} \\
     &\ \psi_{T^2,-,n}^{(\alpha_1,\alpha_{\tau}),M,j}(z) \equiv \psi_{T^2,n-1}^{(\alpha_1,\alpha_{\tau}),M,j}(z) = (-i {\cal D}^{\dagger})^{n-1} \psi_{T^2,0}^{(\alpha_1,\alpha_{\tau}),M,j}(z),
    \label{eq:psiT2-n} \\
    &\begin{array}{l}
         \psi_{T^2,0}^{(\alpha_1,\alpha_{\tau}),M,j}(z) = e^{-\frac{\pi M}{2{\rm Im}\tau}|z|^2} h_{T^2}^{(\alpha_1,\alpha_{\tau}),M}(z),  \\
         h_{T^2}^{(\alpha_1,\alpha_{\tau}),M,j}(z) = (M)^{1/4} e^{2\pi i \frac{j+\alpha_1}{M}\alpha_{\tau}} e^{\frac{\pi M}{2{\rm Im}\tau}z^2} \vartheta
         \begin{bmatrix}
         \frac{j+\alpha_1}{M} \\ -\alpha_{\tau}
         \end{bmatrix}
         (Mz,M\tau),
    \end{array}
    \label{eq:psiT20} \\
    &\ {\cal M}_n^2 = \sum_{k=1}^{n} m_k^2 = \sum_{k=1}^{n} \frac{4\pi M}{{\rm Im}\tau} = \frac{4\pi M}{{\rm Im}\tau}n,
    \label{eq:massT2}
\end{align}
where $j \in \mathbb{Z}/M\mathbb{Z}$ denotes the index of eigenvalue of the shifted operator~\cite{Abe:2014noa} and we use the following commutation relation,
\begin{align}
    [{\cal D}, {\cal D}^{\dagger}] = -4 [D_{\bar{z}},D_{z}] = \frac{4\pi M}{{\rm Im}\tau}.
    \label{eq:DDdaggerrelT2}
\end{align}
The number of the degenerated states is equal to the total magnetic flux on $T^2$, $(2\pi)^{-1} \int_{T^2} F = M$.
We also note that the right hand side of Eq.~(\ref{eq:DDdaggerrelT2}) is equal to $4\pi$ times the flux density.


\subsection{Magnetized $T^2/\mathbb{Z}_N$ orbifold}
\label{subsec:T2ZN}

Next, we review on magnetized $T^2/\mathbb{Z}_N$ orbifold models~\cite{Abe:2013bca}\footnote{See also Refs.~\cite{Abe:2008fi,Abe:2014noa,Kobayashi:2017dyu}.}.
$T^2/\mathbb{Z}_N$ orbifold can be constructed by further identifying the $\mathbb{Z}_N$ twisted point, $\rho z$, with $z$,~i.e. $\rho z \sim z$, where $\rho=e^{2\pi i/N}$ denotes the $\mathbb{Z}_N$ twisted phase.
Note that we can consider $N=2$ with an arbitral modulus $\tau$ and $N=3$, $4$, $6$ with $\tau=\rho$. 
$T^2/\mathbb{Z}_N$ orbifolds have some fixed points, $z_{\rm f.p.}$, for $\mathbb{Z}_N$ twist up to $T^2$ translation,~i.e.,
\begin{align}
    \rho z_{\rm f.p.} + u + v \tau = z_{\rm f.p.}, \quad (\exists u,v \in \mathbb{Z}).
    \label{eq:fixedpoint}
\end{align}
They become singular points of the orbifold with the localized curvature, $\xi^R_{z_{\rm f.p.}}/N = (N-1)/N$.
The spinor wave functions on the magnetized $T^2/\mathbb{Z}_N$ orbifold further satisfy the following $\mathbb{Z}_N$ twisted boundary condition,
\begin{align}
    \begin{array}{l}
        \psi_{T^2/Z_N^m,+,n}^{(\alpha_1,\alpha_{\tau}),M}(\rho z) = \rho^m \psi_{T^2/Z_N^m,+,n}^{(\alpha_1,\alpha_{\tau}),M}(z), \\         \psi_{T^2/Z_N^m,-,n}^{(\alpha_1,\alpha_{\tau}),M}(\rho z) = \rho^{m+1} \psi_{T^2/Z_N^m,-,n}^{(\alpha_1,\alpha_{\tau}),M}(z),
    \end{array}
    \label{eq:BCT2ZN}
\end{align}
in addition to Eq.~(\ref{eq:BCT2}), where $m \in \mathbb{Z}/N\mathbb{Z}$ denotes the $\mathbb{Z}_N$ charge.
Note that we can similarly obtain the following relation related to the $\mathbb{Z}_N$ twist around the fixed point at $z=z_{\rm f.p.}$~\cite{Kobayashi:2022tti}\footnote{See also Refs.~\cite{Abe:2013bca,Sakamoto:2020pev}.},
\begin{align}
    \begin{array}{l}
        \psi_{T^2/Z_N^{m_{\rm f.p.}},+,n}^{(\beta_1,\beta_{\tau}),M}(\rho Z) = \rho^{m_{\rm f.p.}} \psi_{T^2/Z_N^{m_{\rm f.p.}},+,n}^{(\beta_1,\beta_{\tau}),M}(Z), \\         \psi_{T^2/Z_N^{m_{\rm f.p.}},-,n}^{(\beta_1,\beta_{\tau}),M}(\rho Z) = \rho^{m_{\rm f.p.}+1} \psi_{T^2/Z_N^{m_{\rm f.p.}},-,n}^{(\beta_1,\beta_{\tau}),M}(Z),
    \end{array}
    \label{eq:BCT2ZNzfp}
\end{align}
with
\begin{align}
    Z&=z-z_{\rm f.p.}, \label{eq:Z} \\
    (\beta_1,\beta_{\tau})&\equiv(\alpha_1+My^{\rm{f.p.}}_{2},\alpha_{\tau}-My^{\rm{f.p.}}_{1}) \quad({\rm{mod}}\,\,1), \label{eq:SSphasebeta} \\
    m_{{\rm f.p.}} &\equiv N\{u\alpha_1+v\alpha_{\tau}+\tfrac{M}{2}(uv+u y^{\rm{f.p.}}_{2} -v y^{\rm{f.p.}}_{1})\}+m \quad ({\rm{mod}}\,\,N), \label{eq:windingnumber}
\end{align}
where $y^{\rm f.p.}_{i}\ (i=1,2)$ are defined by $z_{\rm f.p.}=y^{\rm f.p.}_{1}+ y^{\rm f.p.}_{2}\tau$.
Hereafter, we focus on the fixed point at $z_{\rm f.p.}=0$.
Hence, wave functions of the spinor on the magnetized $T^2/\mathbb{Z}_N$ orbifold can be written by wave functions on the magnetized $T^2$ as
\begin{align}
    \psi_{T^2/Z_N^m,+,n}^{(\alpha_1,\alpha_{\tau}),M,j}(z) 
    =& {\cal N}_{T^2/Z_N} \sum_{k=0}^{N-1} \rho^{-km} \psi_{T^2,+,n}^{(\alpha_1,\alpha_{\tau}),M,j}(\rho^k z)
    \label{eq:psiT2ZN+n} \\
    =& (-i{\cal D}^{\dagger})^n {\cal N}_{T^2/Z_N} \sum_{k=0}^{N-1} \rho^{-k(m+n)} \psi_{T^2,0}^{(\alpha_1,\alpha_{\tau}),M,j}(\rho^k z)
    \notag \\
    =& (-i{\cal D}^{\dagger})^{n} \psi_{T^2/Z_N^{m+n},0}^{(\alpha_1,\alpha_{\tau}),M,j}(z)
    \notag \\
    \equiv& \psi_{T^2/Z_N^m,n}^{(\alpha_1,\alpha_{\tau}),M,j}(z),
    \notag \\
    \psi_{T^2/Z_N^m,-,n}^{(\alpha_1,\alpha_{\tau}),M,j}(z) 
    =& {\cal N}_{T^2/Z_N} \sum_{k=0}^{N-1} \rho^{-k(m+1)} \psi_{T^2,-,n}^{(\alpha_1,\alpha_{\tau}),M,j}(\rho^k z)
    \label{eq:psiT2ZN-n} \\
    =& (-i{\cal D}^{\dagger})^{n-1} {\cal N}_{T^2/Z_N} \sum_{k=0}^{N-1} \rho^{-k(m+n)} \psi_{T^2,0}^{(\alpha_1,\alpha_{\tau}),M,j}(\rho^k z),
    \notag \\
    =& (-i{\cal D}^{\dagger})^{n-1} \psi_{T^2/Z_N^{m+n},0}^{(\alpha_1,\alpha_{\tau}),M,j}(z)
    \notag \\
    \equiv& \psi_{T^2/Z_N^m,n-1}^{(\alpha_1,\alpha_{\tau}),M,j}(z),
    \notag
\end{align}
where ${\cal N}_{T^2/Z_N}$ denotes the normalization factor and we use the fact that ${\cal D}^{\dagger}$ transforms under the $\mathbb{Z}_N$ twist, $\rho$, as ${\cal D}^{\dagger} \rightarrow \rho^{-1} {\cal D}^{\dagger}$.
Here, we note that the holomorphic part of a wave function with $\mathbb{Z}_N$ charge, $m$, can be expanded by $z^{m+kN}$ with $k \in \mathbb{Z}_+$.
The number of the degenerated states can be determined by the total magnetic flux on $T^2/\mathbb{Z}_N$ orbifold, which includes the localized flux reviewed in the next subsection.


\subsection{Singular gauge transformation}
\label{subsec:singulargauge}

Now, we  remove the $\mathbb{Z}_N$ phase in the $\mathbb{Z}_N$ twisted boundary condition in Eq.~(\ref{eq:BCT2ZN}) by introducing 
the singular gauge transformation~\cite{Kobayashi:2022tti}\footnote{See also Refs.~\cite{Lee:2003mc,Buchmuller:2015eya,Buchmuller:2018lkz,Polchinski:1998rq}.}.
The singular gauge transformation can be defined as
\begin{align}
    \begin{array}{ll}
        A \rightarrow \tilde{A} = A + \delta A, & \delta A = i U_{\xi^F_0} d U_{\xi^F_0}^{-1} = -i \frac{\xi^F_0}{2} \frac{h^{(1)}_1(z)}{h_1(z)} dz + i \frac{\xi^F_0}{2} \frac{\overline{h^{(1)}_1(z)}}{\overline{h_1(z)}} d\bar{z}
        \simeq -i \frac{\xi^F_0}{2} \frac{1}{z} dz + i \frac{\xi^F_0}{2} \frac{1}{\bar{z}} d\bar{z}, \\
        \frac{F}{2\pi} \rightarrow \frac{\tilde{F}}{2\pi} = \frac{F}{2\pi} + \frac{\delta F}{2\pi}, & \frac{\delta F}{2\pi} = i \xi^F_0 \delta(z) \delta(\bar{z}) dz \land d\bar{z},
    \end{array}
    \label{eq:AFtilde}
\end{align}
where $U_{\xi^F_0}$ is defined as
\begin{align}
    U_{\xi^F_0} = \left( \frac{\psi^{(\frac{1}{2},\frac{1}{2}),1}_{T^2/Z_N^1,0}(z)}{\overline{\psi^{(\frac{1}{2},\frac{1}{2}),1}_{T^2/Z_N^1,0}(z)}} \right)^{\frac{\xi^F_0}{2}}= \left( \frac{h_1(z)}{\overline{h_1(z)}} \right)^{\frac{\xi^F_0}{2}}
        \simeq \left( \frac{h^{(1)}_1(0)z}{\overline{h^{(1)}_1(0)z}} \right)^{\frac{\xi^F_0}{2}}, \label{eq:singularU}
\end{align}
with
\begin{align}
    \begin{array}{l}
        \psi^{(\frac{1}{2},\frac{1}{2}),1}_{T^2/Z_N^1,0}(z) =  \psi^{(\frac{1}{2},\frac{1}{2}),1}_{T^2,0}(z) =  e^{-\frac{\pi}{2{\rm Im}\tau}|z|^2} h_1(z) \\
        h_1(z) \equiv h^{(\frac{1}{2},\frac{1}{2}),1}_{T^2,0} = e^{\frac{\pi i}{2}} e^{\frac{\pi}{2{\rm Im}\tau}z^2} \vartheta
        \begin{bmatrix}
        \frac{1}{2} \\ -\frac{1}{2}
        \end{bmatrix}
        (z,\tau)
    \end{array}.
    \label{eq:psia1212M1m1}
\end{align}
It induces the localized flux, $\xi^F_0/N$, at $z_{\rm f.p.}=0$. 
Similarly, the localized curvature at $z_{\rm f.p.}=0$, $\xi^R_0/N = (N-1)/N$, can be introduced by the following gauge transformation for the spin connection,
\begin{align}
    \begin{array}{ll}
        \omega=0 \rightarrow \tilde{\omega} = \omega + \delta \omega = \delta \omega, & \delta \omega = i U_{\xi^R_0} d U_{\xi^R_0}^{-1}, \\
        \frac{R}{2\pi}=0 \rightarrow \frac{\tilde{R}}{2\pi} = \frac{F}{2\pi} + \frac{\delta R}{2\pi} = \frac{\delta R}{2\pi}, & \frac{\delta R}{2\pi} = i \xi^R_0 \delta(z) \delta(\bar{z}) dz \land d\bar{z},
    \end{array}
    \label{eq:wRtilde}
\end{align}
where $U_{\xi^R_0}$ is defined in Eq.~(\ref{eq:singularU}) with replacing $\xi^F_0$ with $\xi^R_0$.
Then, the covariant derivatives are modified as
\begin{align}
    \begin{array}{l}
         D_{z} \rightarrow \tilde{D}_{z} 
         = U_{\xi_0^F} U_{\xi_0^R}^{-s} D_{z} U_{\xi_0^R}^{s} U_{\xi_0^F}^{-1}
         = \partial_{z}-i\tilde{A}_{z}+is\tilde{\omega}_{z}, \\
         D_{\bar{z}} \rightarrow \tilde{D}_{\bar{z}} 
         = U_{\xi_0^F} U_{\xi_0^R}^{-s} D_{\bar{z}} U_{\xi_0^R}^{s} U_{\xi_0^F}^{-1}
         = \partial_{\bar{z}}-i\tilde{A}_{\bar{z}}+is\tilde{\omega}_{\bar{z}},
    \end{array}
    \label{eq:DDbartilde}
\end{align}
where $s$ denotes the spin of the spinor; $s=+1/2$ ($s=-1/2$) for positive (negative) chirality.
Wave functions of the spinor, on the other hand, are transformed by the unitary transformations, $U_{\xi^F_0}$ and $U_{\xi^R_0}$, as
\begin{align}
    \begin{array}{l}
         \tilde{\psi}_{T^2/Z_N^m,+,n}^{(\alpha_1,\alpha_{\tau})}(z) = U_{\xi^F_0} U_{\xi^R_0}^{-1/2} \psi_{T^2/Z_N^m, +,n}^{(\alpha_1,\alpha_{\tau})}(z) 
         = \left( \frac{\psi^{(\frac{1}{2},\frac{1}{2}),1}_{T^2/Z_N^1,0}(z)}{\overline{\psi^{(\frac{1}{2},\frac{1}{2}),1}_{T^2/Z_N^1,0}(z)}} \right)^{\frac{\xi^F_0}{2}-\frac{1}{2}\frac{\xi^R_0}{2}} \psi_{T^2/Z_N^m, +,n}^{(\alpha_1,\alpha_{\tau})}(z),  
         \\
         \tilde{\psi}_{T^2/Z_N^m,-,n}^{(\alpha_1,\alpha_{\tau})}(z) = U_{\xi^F_0} U_{\xi^R_0}^{1/2} \psi_{T^2/Z_N^m, -,n}^{(\alpha_1,\alpha_{\tau})}(z) 
         =  \left( \frac{\psi^{(\frac{1}{2},\frac{1}{2}),1}_{T^2/Z_N^1,0}(z)}{\overline{\psi^{(\frac{1}{2},\frac{1}{2}),1}_{T^2/Z_N^1,0}(z)}} \right)^{\frac{\xi^F_0}{2}+\frac{1}{2}\frac{\xi^R_0}{2}} \psi_{T^2/Z_N^m, -,n}^{(\alpha_1,\alpha_{\tau})}(z). 
    \end{array}
    \label{eq:psitilde}
\end{align}
Accordingly, the $\mathbb{Z}_N$ twisted boundary conditions are modified as
\begin{align}
    \begin{array}{l}
         \tilde{\psi}_{T^2/Z_N^m,+,n}^{(\alpha_1,\alpha_{\tau})}(\rho z) = \rho^{\xi^F_0 - \frac{1}{2}\xi^R_0 + m} \tilde{\psi}_{T^2/Z_N^m,+,n}^{(\alpha_1,\alpha_{\tau})}(z), \\
         \tilde{\psi}_{T^2/Z_N^m,-,n}^{(\alpha_1,\alpha_{\tau})}(\rho z) = \rho^{\xi^F_0 + \frac{1}{2}\xi^R_0 + m+1} \tilde{\psi}_{T^2/Z_N^m,-,n}^{(\alpha_1,\alpha)}(z). 
    \end{array}
    \label{eq:BCmodT2ZN}
\end{align}
To remove the $\mathbb{Z}_N$ phase from the $\mathbb{Z}_N$ twisted boundary condition, we introduce the localized flux,
\begin{align}
    \xi^F_0 = \frac{N-1}{2} - m + \ell N, \label{eq:localizedfluxz0}
\end{align}
at $z=0$, where $\ell \in \mathbb{Z}$ denotes the degree of freedom of the localized flux at $z=0$.
Similarly, from Eq.~(\ref{eq:BCT2ZNzfp}), we can find that the localized flux,
\begin{align}
    \xi^F_{\rm f.p.}=\frac{\xi^R_{\rm f.p.}}{2}-m_{\rm f.p.}+\ell_{\rm f.p.}N, \label{eq:localizedfluxzfp}
\end{align}
is introduced at $z=z_{\rm f.p.}$, where $\xi^R_{\rm f.p.}$ denotes the localized curvature at $z=z_{\rm f.p.}$, $m_{\rm f.p.}$ is defined in Eq.~(\ref{eq:windingnumber}), and $\ell_{\rm f.p.} \in \mathbb{Z}$ denotes the degree of freedom of the localized flux at $z=z_{\rm f.p.}$.
Note that the boundary conditions related to $T^2$ translation in Eq.~(\ref{eq:BCT2}) are also modified as
\begin{align}
    \begin{array}{ll}
        \tilde{\psi}_{T^2/Z_N^m,\pm,n}^{(\alpha_1,\alpha_{\tau})}(z+1) = e^{2\pi i (\alpha_1 + \frac{\xi^F_0}{2}\mp\frac{1}{2}\frac{\xi^R_0}{2})} e^{i \tilde{\chi}_1(z)} \tilde{\psi}_{T^2/Z_N^m,\pm,n}^{(\alpha_1,\alpha_{\tau})}(z), & \tilde{\chi}_1(z) = \pi (M+\xi^F_0\mp\frac{\xi^R_0}{2}) \frac{{\rm Im}z}{{\rm Im}\tau}, \\
        \tilde{\psi}_{T^2/Z_N^m,\pm,n}^{(\alpha_1,\alpha_{\tau})}(z+\tau) =e^{2\pi i (\alpha_{\tau} + \frac{\xi^F_0}{2}\mp\frac{1}{2}\frac{\xi^R_0}{2})} e^{i \tilde{\chi}_{\tau}(z)} \tilde{\psi}_{T^2/Z_N^m,\pm,n}^{(\alpha_1,\alpha_{\tau})}(z), & \tilde{\chi}_{\tau}(z) = \pi (M+\xi^F_0\mp\frac{\xi^R_0}{2}) \frac{{\rm Im}(\bar{\tau}z)}{{\rm Im}\tau}.
    \end{array}
    \label{eq:BCT2mod}
\end{align}
From Eqs.~(\ref{eq:DiraceqT2}) and (\ref{eq:psitilde}), in addition, ${\cal D}$ and ${\cal D}^{\dagger}$ are transformed as
\begin{align}
    \begin{array}{l}
         {\cal D} \rightarrow \tilde{{\cal D}}_{(m,\ell)} = 2 U_{\xi_0^R} \tilde{D}_{\bar{z}}^{(m,\ell)} = 2 \left( \frac{\psi^{(\frac{1}{2},\frac{1}{2}),1}_{T^2/Z_N^1,0}(z)}{\left|\psi^{(\frac{1}{2},\frac{1}{2}),1}_{T^2/Z_N^1,0}(z)\right|} \right)^{N-1} \left( \partial_{\bar{z}} + \frac{\pi M}{2{\rm Im}\tau}z + \frac{\ell N - m}{2\bar{z}} \right),  \\
          {\cal D}^{\dagger} \rightarrow \tilde{{\cal D}}^{\dagger}_{(m,\ell)} = - 2 U_{\xi_0^R}^{-1} \tilde{D}_{z}^{(m+1,\ell+1)} = -2 \left( \frac{\psi^{(\frac{1}{2},\frac{1}{2}),1}_{T^2/Z_N^1,0}(z)}{\left|\psi^{(\frac{1}{2},\frac{1}{2}),1}_{T^2/Z_N^1,0}(z)\right|} \right)^{-(N-1)} \left( \partial_{z} - \frac{\pi M}{2{\rm Im}\tau}\bar{z} - \frac{(\ell+1)N-(m+1)}{2z} \right).
    \end{array}
    \label{eq:calDDdaggertilde}
\end{align}
It means that the gamma matrices are also affected by the singular gauge transformation.
Hence, by using the singular gauge transformed wave functions, Eqs.~(\ref{eq:psiT2ZN+n}) and (\ref{eq:psiT2ZN-n}) can be modified as
\begin{align}
    \tilde{\psi}_{T^2/Z_N^{(m,\ell)},+,n}^{(\alpha_1,\alpha_{\tau}),M,j}(z) 
    =& (-i\tilde{{\cal D}}_{(m,\ell)}^{\dagger})(-i\tilde{{\cal D}}_{(m+1,\ell+1)}^{\dagger}) \cdots (-i\tilde{{\cal D}}_{(m+n-1,\ell+n-1)}^{\dagger}) \tilde{\psi}_{T^2/Z_N^{(m+n,\ell+n)},0}^{(\alpha_1,\alpha_{\tau}),M,j}(z)
    \notag \\
    \equiv& \prod_{k=1}^{n} (-i\tilde{{\cal D}}_{(m+k-1,\ell+k-1)}^{\dagger}) \tilde{\psi}_{T^2/Z_N^{(m+n,\ell+n)},0}^{(\alpha_1,\alpha_{\tau}),M,j}(z)
    \label{eq:psiT2ZN+ntilde} \\
    \equiv& \tilde{\psi}_{T^2/Z_N^{(m,\ell)},n}^{(\alpha_1,\alpha_{\tau}),M,j}(z),
    \notag \\
    \tilde{\psi}_{T^2/Z_N^{(m,\ell)},-,n}^{(\alpha_1,\alpha_{\tau}),M,j}(z) 
    =& (-i\tilde{{\cal D}}_{(m+1,\ell+1)}^{\dagger})(-i\tilde{{\cal D}}_{(m+2,\ell+2)}^{\dagger}) \cdots (-i\tilde{{\cal D}}_{(m+n-1,\ell+n-1)}^{\dagger}) \tilde{\psi}_{T^2/Z_N^{(m+n,\ell+n)},0}^{(\alpha_1,\alpha_{\tau}),M,j}(z)
    \notag \\
    \equiv& \prod_{k=2}^{n} (-i\tilde{{\cal D}}_{(m+k-1,\ell+k-1)}^{\dagger}) \tilde{\psi}_{T^2/Z_N^{(m+n,\ell+n)},0}^{(\alpha_1,\alpha_{\tau}),M,j}(z)
    \label{eq:psiT2ZN-ntilde} \\
    \equiv& \tilde{\psi}_{T^2/Z_N^{(m+1,\ell+1)},n-1}^{(\alpha_1,\alpha_{\tau}),M,j}(z).
    \notag
\end{align}
Note that the holomorphic part of a wave function with $\mathbb{Z}_N$ charge, $m$, after the singular gauge transformation can be expanded by $z^{(\ell+k)N}$ with $k \in \mathbb{Z}_+$.
The number of the degenerated states of $\tilde{\psi}_{T^2/Z_N^{(m,\ell)},n}^{(\alpha_1,\alpha_{\tau}),M,j}$ is the same as that of zero modes, $\tilde{\psi}_{T^2/Z_N^{(m+n,\ell+n)},0}^{(\alpha_1,\alpha_{\tau}),M,j}$.
The number of zero modes, $\tilde{\psi}_{T^2/Z_N^{(m,\ell)},0}^{(\alpha_1,\alpha_{\tau}),M,j}$ can be determined by the total magnetic flux on $T^2/\mathbb{Z}_N$ orbifold,
\begin{align}
    \int_{T^2/\mathbb{Z}_N} \frac{\tilde{F}}{2\pi}=\frac{M}{N}+\sum_{\rm f.p.}\frac{\xi^F_{\rm f.p.}}{N}=\left( \frac{M}{N}-\sum_{\rm f.p.}\frac{m_{\rm f.p.}}{N}+1 \right)+\sum_{\rm f.p.}\ell_{\rm f.p.}, \label{eq:degeneratenumber}
\end{align}
where the first term (in the parentheses) shows the number of bulk modes and the second term, $\ell_{\rm f.p.}$, shows the number of the localized modes at $z=z_{\rm f.p.}$, discussed in section~\ref{sec:localizedmode}.
Then, we can obtain the number of $\tilde{\psi}_{T^2/Z_N^{(m+n,\ell+n)},0}^{(\alpha_1,\alpha_{\tau}),M,j}$ as well as $\tilde{\psi}_{T^2/Z_N^{(m,\ell)},n}^{(\alpha_1,\alpha_{\tau}),M,j}$ by replacing the $\mathbb{Z}_N$ charge and the localized flux from $(m,\ell)$ to $(m+n,\ell+n)$, respectively.
Furthermore, since the commutation relation in Eq.~(\ref{eq:DDdaggerrelT2}) is modified as
\begin{align}
    -4[\tilde{D}_{\bar{z}}^{(m+k,\ell+k)}, \tilde{D}_{z}^{(m+k,\ell+k)}] = 4N \left( \frac{\pi M}{N{\rm Im}\tau} + \frac{(\ell+k)N-(m+k)}{N} \delta(z) \right),
    \label{calDDdaggertilderel}
\end{align}
due to Eq.~(\ref{eq:calDDdaggertilde}), the mass squared eigenvalue around $z=0$, ${\cal M}_n^2(z)$, is also modified as
\begin{align}
    \tilde{{\cal M}}_n^2(z) 
    =& \sum_{k=1}^{n} \tilde{m}_k^2 \notag \\
    =& \sum_{k=1}^{n} 4N \left( \frac{\pi M}{N{\rm Im}\tau} + \frac{(\ell +k)N -(m+k)}{N} \delta(z) \right) \notag \\
    =& 4N \left( \frac{\pi M}{N{\rm Im}\tau} + \left( \frac{\ell N -m}{N} + \frac{N-1}{2N}(n+1) \right) \delta(z) \right)n.
    \label{eq:masstilde}
\end{align}
By considering the other fixed points, the mass squared eigenvalue, ${\cal M}_n^2(z)$, is written by
\begin{align}
    \tilde{{\cal M}}_n^2(z)
    = 4N \left( \frac{\pi M}{N{\rm Im}\tau} + \sum_{\rm f.p.} \left( \frac{\ell_{\rm f.p.} N -m_{\rm f.p.}}{N} + \frac{\xi^R_{\rm f.p.}}{2N}(n+1) \right) \delta(z-z_{\rm f.p.}) \right)n.
    \label{eq:masstildetotal}
\end{align}
Here, this $\tilde{{\cal M}}_n^2(z)$ is the  mass squared eigenvalue on the compact space, $T^2/\mathbb{Z}_N$ orbifold.
The physical mass squared on the four-dimensional space-time, $\tilde{{\cal M}}_{4D,n}^2$, can be obtained by the following overlap integral on $T^2/\mathbb{Z}_N$ orbifold,
\begin{align}
    &\tilde{{\cal M}}_{4D,n,ij}^2 \notag \\
    =& \int dzd\bar{z} \tilde{{\cal M}}_n^2(z) 
    \left( \tilde{\psi}_{T^2/Z_N^{(m,\ell)},n}^{(\alpha_1,\alpha_{\tau}),M,i}(z) \right)^{\dagger} \tilde{\psi}_{T^2/Z_N^{(m,\ell)},n}^{(\alpha_1,\alpha_{\tau}),M,j}(z) \notag \\
    =& 4N \left( \frac{\pi M}{N{\rm Im}\tau} \delta_{i,j} + \sum_{\rm f.p.} \left( \frac{\ell_{\rm f.p.} N -m_{\rm f.p.}}{N} + \frac{\xi^{R}_{{\rm f.p.}}}{2N}(n+1) \right) \left( \psi_{T^2/Z_N^m,n}^{(\alpha_1,\alpha_{\tau}),M,i}(z_{\rm f.p.}) \right)^{\ast} \left( \psi_{T^2/Z_N^m,n}^{(\alpha_1,\alpha_{\tau}),M,j}(z_{\rm f.p.}) \right) \right)n
    \notag \\
    =& 4N \left( \frac{\pi M}{N{\rm Im}\tau} \delta_{i,j} + \sum_{\rm f.p.} \left( \frac{\ell_{\rm f.p.} N -m_{\rm f.p.}}{N} + \frac{\xi^{R}_{{\rm f.p.}}}{2N}(n+1) \right) \left( \psi_{T^2/Z_N^{m_{\rm f.p.}},n}^{(\beta_1,\beta_{\tau}),M,i}(0) \right)^{\ast} \left( \psi_{T^2/Z_N^{m_{\rm f.p.}},n}^{(\beta_1,\beta_{\tau}),M,j}(0) \right) \right)n.
    \label{eq:tildecalM4Dnij}
\end{align}
Note that $\psi_{T^2/Z_N^{m_{\rm f.p.}},n}^{(\beta_1,\beta_{\tau}),M,j}(0)=0$ when $m_{\rm f.p.} \neq 0$.
By the basis transformation, $\psi_{T^2/Z_N^m,n}^{(\alpha_1,\alpha_{\tau}),M,j'}(z) = U_{j'j}(z) \psi_{T^2/Z_N^m,n}^{(\alpha_1,\alpha_{\tau}),M,j}(z)$, such that
\begin{align}
    \begin{array}{l}
        U(z_{\rm f.p.}) = \prod_{J} \left( U^{J(J+1)} \right) {\rm diag}\left(e^{-i{\rm arg}\left(\psi_{T^2/Z_N^m,n}^{(\alpha_1,\alpha_{\tau}),M,j}(z_{\rm f.p.}) \right)}\right), \\
        U^{J(J+1)} =
        \begin{pmatrix}
        1 & \ & \ & \ & \ \\
        \ & \ddots & \ & \ & \ \\
        \ & \ &\begin{array}{cc}
        \cos \theta_{J(J+1)} & - \sin \theta_{J(J+1)} \\
        \sin \theta_{J(J+1)} & \cos \theta_{J(J+1)}
        \end{array}& \ & \ \\
        \ & \ & \ & \ddots & \ \\
        \ & \ & \ & \ & \ & 1
        \end{pmatrix}, \\
        \tan^2 \theta_{J(J+1)} = \frac{\sum_{I=1}^{J} \left|\psi_{T^2/Z_N^m,n}^{(\alpha_1,\alpha_{\tau}),M,I}(z_{\rm f.p.})\right|^2}{\left|\psi_{T^2/Z_N^m,n}^{(\alpha_1,\alpha_{\tau}),M,J+1}(z_{\rm f.p.})\right|^2} =
        \frac{\sum_{I=1}^{J} \left|\psi_{T^2/Z_N^{m_{\rm f.p.}},n}^{(\beta_1,\beta_{\tau}),M,I}(0)\right|^2}{\left|\psi_{T^2/Z_N^{m_{\rm f.p.}},n}^{(\beta_1,\beta_{\tau}),M,J+1}(0)\right|^2},
    \end{array} \label{eq:unitary}
\end{align}
Eq.~(\ref{eq:tildecalM4Dnij}) can be diagonalized as
\begin{align}
    &\tilde{{\cal M}}_{4D,n,i'j'}^2 \notag \\
    =& 4N \left( \frac{\pi M}{N{\rm Im}\tau} \delta_{i',j'} + \sum_{\rm f.p.} \left[ \left( \sum_{j} \left| \psi_{T^2/Z_N^{m_{\rm f.p.}},n}^{(\beta_1,\beta_{\tau}),M,j}(0) \right|^2 \right) \left( \frac{\ell_{\rm f.p.} N -m_{\rm f.p.}}{N} + \frac{\xi^R_{\rm f.p.}}{2N}(n+1) \right) \right] \delta_{i',j'_{{\rm max}}} \right)n.
    \label{eq:tildecalM4Dn}
\end{align}
Hence, when $\exists m_{\rm f.p.}=0$, only the physical mass squared of $j'={j'}_{\rm max}$ mode is modified from Eq.~(\ref{eq:massT2}).

We also comment about the modular weight of singular gauge transformed wave functions on magnetized $T^2/\mathbb{Z}_2$ orbifold.
When we regard the wave functions as the periodic functions defined in Ref.~\cite{Jeric:2025iwk},\footnote{See for periodic functions of modular forms Ref.~\cite{Ding:2020zxw}.} the unitary transformation, $\left( \psi^{(\frac{1}{2},\frac{1}{2}),1}_{T^2/Z_N^1,0}(z)/\left|\psi^{(\frac{1}{2},\frac{1}{2}),1}_{T^2/Z_N^1,0}(z)\right| \right)$, is proportional to $e^{-i\theta/2}$.
Hence, the singular gauge transformed wave function of level $n$ on magnetized $T^2/\mathbb{Z}_N$ orbifold is proportional to $e^{-i\left(n+(\xi^F_0-s\xi^R_0+1)/2\right)\theta}=e^{-i(n+(1-m)/2+\ell)\theta}$, and then the modular weight becomes $n+(\xi^F_0-s\xi^R_0+1)/2=n+(1-m)/2+\ell$ since it is an eigenvalue of $\hat{H}=-i\partial_{\theta}$, defined in Ref.~\cite{Jeric:2025iwk}.
This is consistent with analysis in Ref.~\cite{Kikuchi:2023clx}. 
It is interesting that we can realize larger modular weights by the localized flux $\ell$ even for the massless modes $n=0$.
That is useful in modular flavor models \cite{Feruglio:2017spp,Kobayashi:2018vbk,Penedo:2018nmg,Novichkov:2018nkm,Kobayashi:2018scp}. (See for reviews \cite{Kobayashi:2023zzc,Ding:2023htn}.)


\section{Magnetized $S^2$}
\label{sec:S2}

In this section, we review on magnetized $S^2$ models.
When we denote the coordinate of $S^2 \simeq \mathbb{CP}^1$ as $z'$, the metric of $S^2$ is written by
\begin{align}
    ds^2 = 2{h'}_{z'\bar{z}'} dz' d\bar{z}' = \frac{R^4}{(R^2+|z'|^2)} dz' d\bar{z}',
    \label{eq:metricS2}
\end{align}
where $R$ denotes the radius of $S^2$, and then the area of $S^2$ is $4\pi R^2$.
The gamma matrices, $\gamma^{z'}$ and $\gamma^{\bar{z}'}$, are given by 
\begin{align}
    \gamma^{z'} =
    \begin{pmatrix}
        0 & \frac{R^2+|z'|^2}{R^2} \\
        0 & 0
    \end{pmatrix},
    \quad
    \gamma^{\bar{z}'} =
    \begin{pmatrix}
        0 & 0 \\
        \frac{R^2+|z'|^2}{R^2} & 0
    \end{pmatrix},
    \label{eq:gammaprime}
\end{align}
which satisfy $\{\gamma^{z'},\gamma^{\bar{z}'}\}=2{h'}^{z'\bar{z}'}$.
Here, ${h'}^{z'\bar{z}'}$ denotes the inverse of ${h'}_{z'\bar{z}'}$.
The total curvature of $S^2$ is $(2\pi)^{-1} \int_{S^2} R' = \chi(S^2) = 2$.
The spin connection is given by
\begin{align}
    \omega' = \Sigma^{12} (\omega'_{z'12}dz' + \omega'_{\bar{z'}12} d\bar{z}') = \frac{i}{2}\sigma_{3} \left( -\frac{i}{2} \frac{2}{R^2 + |z'|^2}\bar{z}' dz' + \frac{i}{2} \frac{2}{R^2 + |z'|^2}z' d\bar{z}' \right).
    \label{eq:wprime}
\end{align}


\subsection{Magnetic flux background}
\label{subsec:withoutvortex}

On the $S^2$, the $U(1)$ magnetic flux,
\begin{align}
    \frac{F'}{2\pi} = \frac{i}{2\pi} \frac{R^2 M'}{(R^2 + |z'|^2)^2} dz' \land d\bar{z}',
    \label{eq:Fprime}
\end{align}
is inserted~\cite{Conlon:2008qi}, where it satisfies $(2\pi)^{-1} \int_{S^2} F' = M'$.
Hereafter, we consider $M'>0$.
The magnetic flux is induced by the gauge potential,
\begin{align}
     A' = -\frac{i}{2} \frac{M'}{R^2 + |z'|^2}\bar{z}' dz' +  \frac{i}{2} \frac{M'}{R^2 + |z'|^2}\bar{z}' d\bar{z}'.
     \label{eq:Aprime}
\end{align}
The covariant derivatives are written by ${D'}_{z'}=\partial_{z'}-iqA_{z'}+\frac{i}{2}\sigma_{3}\omega'_{z'12}$ and ${D'}_{\bar{z}'}=\partial_{\bar{z}'}-iqA_{\bar{z}'}+\frac{i}{2}\sigma_{3}\omega'_{\bar{z}'12}$, where $q$ denotes the $U(1)$ charge of a field.
In the following, we consider a two-dimensional spinor, ${\psi'}_{S^2}^{M'}(z')={^t}({\psi'}_{S^2,+}^{M'}(z'), {\psi'}_{S^2,-}^{M'}(z'))$, on the magnetized $S^2$ with $U(1)$ unit charge $q=1$.
The spinor satisfies the following Dirac equation,
\begin{align}
    \begin{array}{l}
        -i {{\cal D}}^{\prime\dagger}_{(M')} {\psi}_{S^2,-,n}^{\prime M'}(z) = {\psi}_{S^2,+,n}^{\prime M'}(z'), \\
        i {{\cal D}^{\prime}}_{(M')} {\psi}_{S^2,+,n}^{\prime M'}(z') = {{\cal M}^{\prime}}_n^2 {\psi}_{S^2,-,n}^{\prime M'}(z'),
    \end{array}
    \label{eq:DiraceqS2}
\end{align}
where ${{\cal D}'}$ and ${{\cal D}'}^{\dagger}$ are defined from the Dirac operator,
\begin{align}
    i \slashed{D}' \equiv
    \begin{pmatrix}
        0 & -i{{\cal D}}^{\prime\dagger}_{(M')} \\
        i{{\cal D}^{\prime}}_{(M')} & 0
    \end{pmatrix}
     =&
    \begin{pmatrix}
        0 & \frac{R^2+|z'|^2}{R^2}i{D}_{z'}^{\prime(M'+1)} \\
        \frac{R^2+|z'|^2}{R^2}i{D}_{\bar{z}'}^{\prime(M'-1)} & 0
    \end{pmatrix}
    \notag \\
    =&
    \begin{pmatrix}
        0 & \frac{R^2+|z'|^2}{R^2}i(\partial_{z'}-\frac{M'+1}{2(R^2+|z'|^2)}\bar{z}') \\
        \frac{R^2+|z'|^2}{R^2}i(\partial_{\bar{z}'}+\frac{M'-1}{2(R^2+|z'|^2)}z') & 0
    \end{pmatrix},
    \label{eq:DiracopS2}
\end{align}
and ${{\cal M}^{\prime}}_{n}$ denotes the mass eigenvalue of the spinor with level $n$.\footnote{In this paper, we define the spinor such that the mass eigenvalue only appears in the second equation in Eq.~(\ref{eq:DiraceqS2}).}
The solutions of the Dirac equation in Eq.~(\ref{eq:DiraceqS2}) are obtained as
\begin{align}
    {\psi}_{S^2,+,n}^{\prime M',a'}(z') 
    =& (-i{{\cal D}}_{(M')}^{\prime\dagger})(-i{{\cal D}}_{(M'+2)}^{\prime\dagger}) \cdots (-i{{\cal D}}_{(M'+2(n-1))}^{\prime\dagger}) {\psi}_{S^2,0}^{\prime M'+2n,a'+n}(z')
    \notag \\
    \equiv& \prod_{k=1}^{n} (-i{{\cal D}}_{(M'+2(k-1))}^{\prime\dagger}) {\psi}_{S^2,0}^{\prime M'+2n,a'+n}(z')
    \label{eq:psiS2+n} \\
    \equiv& {\psi}_{S^2,n}^{\prime M',a'}(z'),
    \notag \\
    {\psi}_{S^2,-,n}^{\prime M',a'}(z') 
    =& (-i{{\cal D}}_{(M'+2)}^{\prime\dagger})(-i{{\cal D}}_{(M'+4)}^{\prime\dagger}) \cdots (-i{{\cal D}}_{(M'+2(n-1))}^{\prime\dagger}) {\psi}_{S^2,0}^{\prime M'+2n,a'+n}(z')
    \notag \\
    \equiv& \prod_{k=2}^{n} (-i{{\cal D}}_{(M'+2(k-1))}^{\prime\dagger}) {\psi}_{S^2,0}^{\prime M'+2n,a'+n}(z')
    \label{eq:psiS2-n} \\
    \equiv& {\psi}_{S^2,n-1}^{\prime M'+2,a'+1}(z'),
    \notag \\
         {\psi}_{S^2,0}^{\prime M',a'}(z') =& \left( \frac{R^2}{R^2+|z'|^2} \right)^{\frac{M'-1}{2}} {h}_{S^2}^{\prime M',a'}(z'), \label{eq:psiS20} \\
         {h}_{S^2}^{\prime M',a'}(z') =& {C}^{\prime(M',a')} \left( \frac{z'}{R} \right)^{a'}, \notag \\
    {{\cal M}}_n^{\prime 2} =& \sum_{k=1}^{n} {m}_k^{\prime 2} = \sum_{k=1}^{n} \frac{M'+2k-1}{R^2} = \frac{M'+n}{R^2}n,
    \label{eq:massS2}
\end{align}
where $(a'+n) \in \mathbb{Z}/(M'+2n)\mathbb{Z}$ denotes the index of eigenvalue of the angular momentum operator, $J^{\prime 3}_{n}$, defined in subsection \ref{subsec:AM}, and we use the following commutation relation,
\begin{align}
    - \left( \frac{R^2+|z'|^2}{R^2} \right) ^2 [D_{\bar{z}'}^{(M'+2k-1)},D_{z'}^{(M'+2k-1)}] = \frac{M'+2k-1}{R^2}.
    \label{eq:DDdaggerrelS2}
\end{align}
The number of the degenerated states of ${\psi}_{S^2,n}^{\prime M',a'}$ is the same as the number of zero modes, ${\psi}_{S^2,0}^{\prime M'+2n,a'+n}$.
The number of zero modes, ${\psi}_{S^2,0}^{\prime M',a'}$, is equal to the total magnetic flux on $S^2$, $(2\pi)^{-1} \int_{S^2} F'= M'$.
Then, we can obtain the number of ${\psi}_{S^2,0}^{\prime M'+2n,a'+n}$ as well as the number of ${\psi}_{S^2,n}^{\prime M',a'}$ by replacing $M'$ with $M'+2n$.
We also note that the right hand side of Eq.~(\ref{eq:DDdaggerrelS2}) is equal to $4\pi$ times the effective magnetic flux density.


\subsection{Vortex background}
\label{subsec:withvortex}

Here, we consider that the vortex, $v'$, is also inserted at $z'=0$~\cite{Dolan:2020sjq} in addition to the $U(1)$ magnetic flux in Eq.~(\ref{eq:Fprime})~i.e.,
\begin{align}
    \frac{\tilde{F}'}{2\pi} = \frac{F'}{2\pi} + \frac{\delta F'}{2\pi}, \quad \frac{\delta F'}{2\pi} = i v' \delta(z') \delta(\bar{z}') dz' \land d\bar{z}'.
    \label{eq:Ftildeprime}
\end{align}
The vortex is induced by the following vector potential,
\begin{align}
    \delta A' = -i \frac{v'}{2} \frac{1}{z'} dz' + i \frac{v'}{2} \frac{1}{\bar{z}'} d\bar{z}',
    \label{eq:deltaAprime}
\end{align}
and then the total vector potential is written by
\begin{align}
    \tilde{A}' = A' + \delta A'.
    \label{eq:Atildeprime}
\end{align}
In this case, the covariant derivatives and also the Dirac operator are modified as
\begin{align}
    D^{\prime}_{z'} \rightarrow \tilde{D}^{\prime}_{z'} = \partial_{z'}-iq\tilde{A}^{\prime}_{z'}+\frac{i}{2}\sigma_{3}\omega^{\prime}_{z'12}, \quad
    D^{\prime}_{\bar{z}'} \rightarrow \tilde{D}^{\prime}_{\bar{z}'} = \partial_{\bar{z}'}-iq\tilde{A}^{\prime}_{\bar{z}'}+\frac{i}{2}\sigma_{3}\omega^{\prime}_{\bar{z}'12},
    \label{eq:covariantderivativetilde}
\end{align}
\begin{align}
    i {\slashed{D}}' \rightarrow  i {\slashed{\tilde{D}}}'  &\equiv    \begin{pmatrix}
        0 & -i{{\tilde{\cal D}}}^{\prime\dagger}_{({M'})} \\
        i{{\tilde{\cal D}}^{\prime}}_{({M'})} & 0
    \end{pmatrix}
    \notag \\
     =&
    \begin{pmatrix}
        0 & \frac{R^2+|{z'}|^2}{R^2}i{\tilde{D}}_{z'}^{\prime(M'+1)} \\
        \frac{R^2+|{z'}|^2}{R^2}i\tilde{D}_{\bar{z}'}^{\prime(M'-1)} & 0
    \end{pmatrix}
    \notag \\
    =&
    \begin{pmatrix}
        0 & \frac{R^2+|{z'}|^2}{R^2}i(\partial_{z'}-\frac{M'+1}{2(R^2+|{z'}|^2)}\bar{z}'-\frac{v'}{2z'}) \\
        \frac{R^2+|{z'}|^2}{R^2}i(\partial_{\bar{z}'}+\frac{M'-1}{2(R^2+|{z'}|^2)}z'+\frac{v'}{2\bar{z}'}) & 0
    \end{pmatrix}.
    \label{eq:DiracoptildeS2}
\end{align}
Then, the solutions of the modified Dirac equation,
\begin{align}
    \begin{array}{l}
        -i {\tilde{{\cal D}}}^{\prime\dagger}_{(M')} {\tilde{\psi}}_{S^2,-,n}^{\prime M'}(z) = {\tilde{\psi}}_{S^2,+,n}^{\prime M'}(z'), \\
        i {\tilde{{\cal D}}^{\prime}}_{(M')} {\tilde{\psi}}_{S^2,+,n}^{\prime M'}(z') = {\tilde{{\cal M}}}_n^{\prime 2}(z') {\tilde{\psi}}_{S^2,-,n}^{\prime M'}(z'),
    \end{array}
    \label{eq:DiraceqmodiS2}
\end{align}
are obtained as
\begin{align}
    {\tilde{\psi}}_{S^2,+,n}^{\prime M',a'}(z') 
    =& (-i{\tilde{{\cal D}}}_{(M')}^{\prime\dagger})(-i{\tilde{{\cal D}}}_{(M'+2)}^{\prime\dagger}) \cdots (-i{\tilde{{\cal D}}}_{(M'+2(n-1))}^{\prime\dagger}) {\tilde{\psi}}_{S^2,0}^{\prime M'+2n,a'+n}(z')
    \notag \\
    \equiv& \prod_{k=1}^{n} (-i{\tilde{{\cal D}}}_{(M'+2(k-1))}^{\prime\dagger}) {\tilde{\psi}}_{S^2,0}^{\prime M'+2n,a'+n}(z')
    \label{eq:psitildeS2+n} \\
    \equiv& {\tilde{\psi}}_{S^2,n}^{\prime M',a'}(z'),
    \notag \\
    {\tilde{\psi}}_{S^2,-,n}^{\prime M',a'}(z') 
    =& (-i{\tilde{{\cal D}}}_{(M'+2)}^{\prime\dagger})(-i{\tilde{{\cal D}}}_{(M'+4)}^{\prime\dagger}) \cdots (-i{\tilde{{\cal D}}}_{(M'+2(n-1))}^{\prime\dagger}) {\tilde{\psi}}_{S^2,0}^{\prime M'+2n,a'+n}(z')
    \notag \\
    \equiv& \prod_{k=2}^{n} (-i{\tilde{{\cal D}}}_{(M'+2(k-1))}^{\prime\dagger}) {\tilde{\psi}}_{S^2,0}^{\prime M'+2n,a'+1}(z')
    \label{eq:psitildeS2-n} \\
    \equiv& {\tilde{\psi}}_{S^2,n-1}^{\prime M'+2,a'+1}(z'),
    \notag \\
         {\tilde{\psi}}_{S^2,0}^{\prime M',a'}(z') =& \left( \frac{R^2}{R^2+|z'|^2} \right)^{\frac{M'-1}{2}} \left( \frac{|z'|}{R} \right)^{-v'}{h}_{S^2}^{\prime M',a'}(z'), \label{eq:psitildeS20} \\
         {h}_{S^2}^{\prime M',a'}(z') =&  {C}^{\prime(M',a')} \left( \frac{z'}{R} \right)^{a'} \quad (a' \in \mathbb{Z}/(M'+v')\mathbb{Z}), \notag \\
    {\tilde{{\cal M}}}_n^{\prime 2}(z') =& \sum_{k=1}^{n} {\tilde{m}}_k^{\prime 2} = \sum_{k=1}^{n} \left(\frac{M'+2k-1}{R^2}+v'\delta(z')\right) ,
    \label{eq:masstildeS2}
\end{align}
where we use the following commutation relation,
\begin{align}
    - \left( \frac{R^2+|z'|^2}{R^2} \right) ^2 [\tilde{D}_{\bar{z}'}^{\prime(M'+2k-1)},\tilde{D}_{z'}^{\prime(M'+2k-1)}] = \frac{M'+2k-1}{R^2} + v'\delta(z').
    \label{eq:DDdaggerrelmodiS2}
\end{align}
The number of zero modes is equal to the total magnetic flux including the vortex on $S^2$.

\subsection{Angular momentum operator}
\label{subsec:AM}

Here, we comment on the angular momentum  operators on magnetized $S^2$~\cite{Dolan:2020sjq}.
They are defined as
\begin{align}
    J^{\prime +}_{k}
    =& \frac{1}{R} \left( z'^2 \tilde{D}^{\prime(M'+2k-1)}_{z'} + R^2 \tilde{D}^{\prime(M'+2k-1)}_{\bar{z}'} - R^2 \left( \frac{M'+2k-1}{R^2+|z'|^2} z' + \frac{v'}{\bar{z}'} \right) \right),
    \label{eq:Jprime+} \\
    J^{\prime -}_{k}
    =& - \frac{1}{R} \left( \bar{z}'^2 \tilde{D}^{\prime(M'+2k-1)}_{\bar{z}'} + R^2 \tilde{D}^{\prime(M'+2k-1)}_{z'} + R^2 \left( \frac{M'+2k-1}{R^2+|z'|^2} \bar{z}' + \frac{v'}{z'} \right) \right), \label{eq:Jprime-} \\
    J^{\prime 3}_{k}
    =& z'\partial_{z'} - \bar{z}'\partial_{\bar{z}'} - \frac{M'+2k-1}{2},
    \label{eq:Jprime3}
\end{align}
and they satisfy the following algebraic relations,
\begin{align}
    &[ J^{\prime +}_{k_+}, J^{\prime -}_{k_-} ] = 2 J^{\prime 3}_{\frac{k_+ + k_-}{2}}, \label{eq:primeJ+J-} \\
    &[ J^{\prime 3}_{k_3}, J^{\prime +}_{k_+} ] = + J^{\prime +}_{k_+}, \label{eq:primeJ3J+} \\
    &[ J^{\prime 3}_{k_3}, J^{\prime -}_{k_-} ] = - J^{\prime -}_{k_-}.
    \label{eq:primeJ3J-}
\end{align}
In addition, it satisfies that
\begin{align}
    \begin{array}{l}
        J^{\prime 3}_{0} {\tilde{\psi}}_{S^2,n}^{\prime M',a'}(z') = \left(a'-\frac{M'-1}{2}\right) {\tilde{\psi}}_{S^2,n}^{\prime M',a'}(z'),  \\
        J^{\prime 3}_{n} {\tilde{\psi}}_{S^2,0}^{\prime M'+2n,a'+n}(z') = \left(a'-\frac{M'-1}{2}\right) {\tilde{\psi}}_{S^2,0}^{\prime M'+2n,a'+n}(z').
    \end{array}
    \label{eq:Jprime3eigenvalue}
\end{align}
Hence, $a'$ means a degree of freedom of the angular momentum and $J^{\prime +}_{k}$ ($J^{\prime -}_{k}$) raises (lowers) the angular momentum.

In section~\ref{sec:localizedmode}, by defining the angular momentum operators on magnetized $T^2/\mathbb{Z}_N$ orbifold from those on magnetized $S^2$ in Eqs.~(\ref{eq:Jprime+})-(\ref{eq:Jprime3}) through the blow-up procedure, we study massive modes of localized modes.


\section{Magnetized blow-up manifold of $T^2/\mathbb{Z}_N$ orbifold}
\label{sec:blowupmanifold}

In this section, let us consider a spinor on the magnetized blow-up manifold of $T^2\mathbb{Z}_N$ orbifold.

\begin{figure}[H]
    \centering
    \begin{minipage}{7cm}
    \centering
    \includegraphics[bb=0 0 750 700,width=5cm]{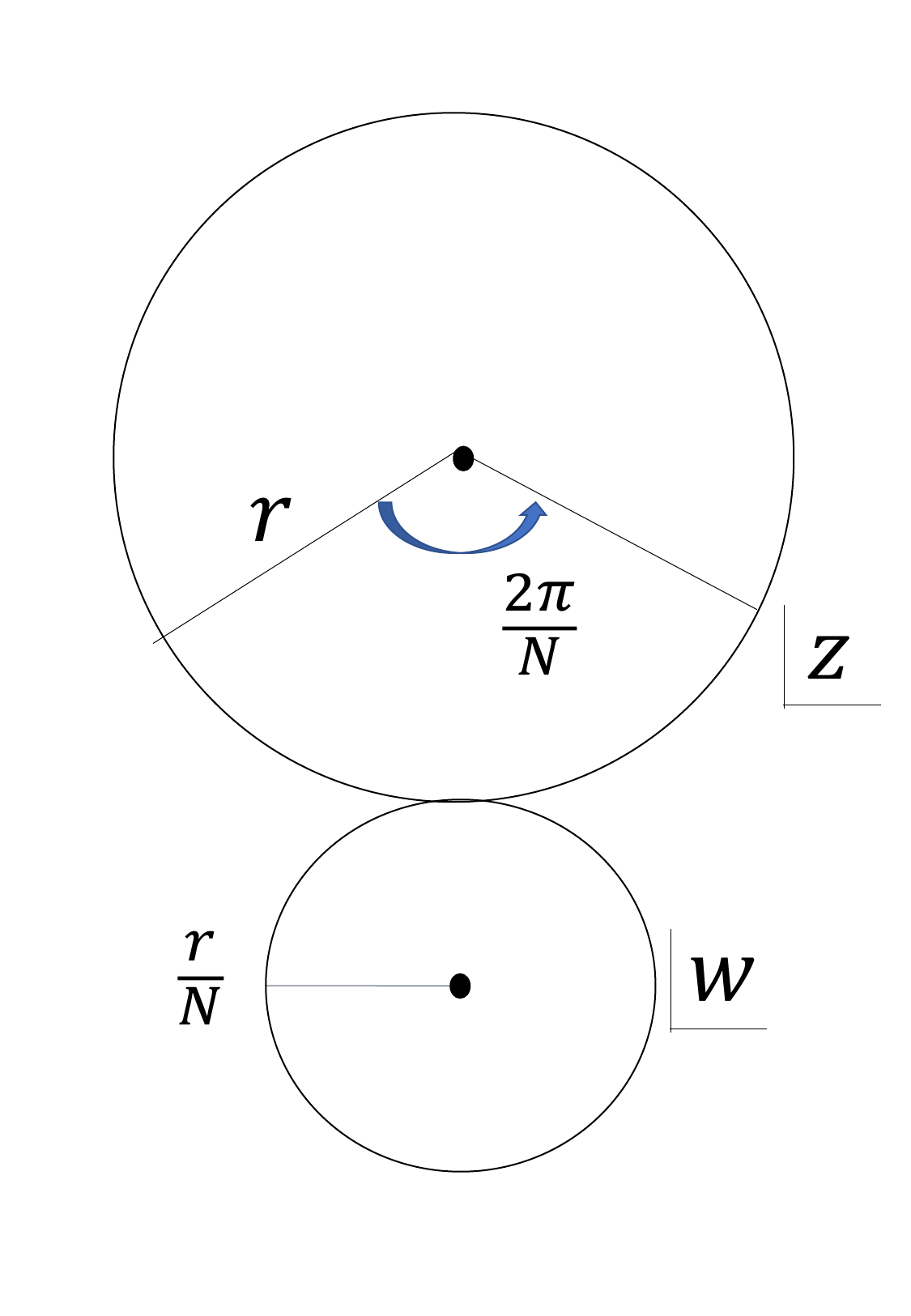}
    \end{minipage}
    \begin{minipage}{7cm}
    \centering
    \includegraphics[bb=0 0 500 375,width=5cm]{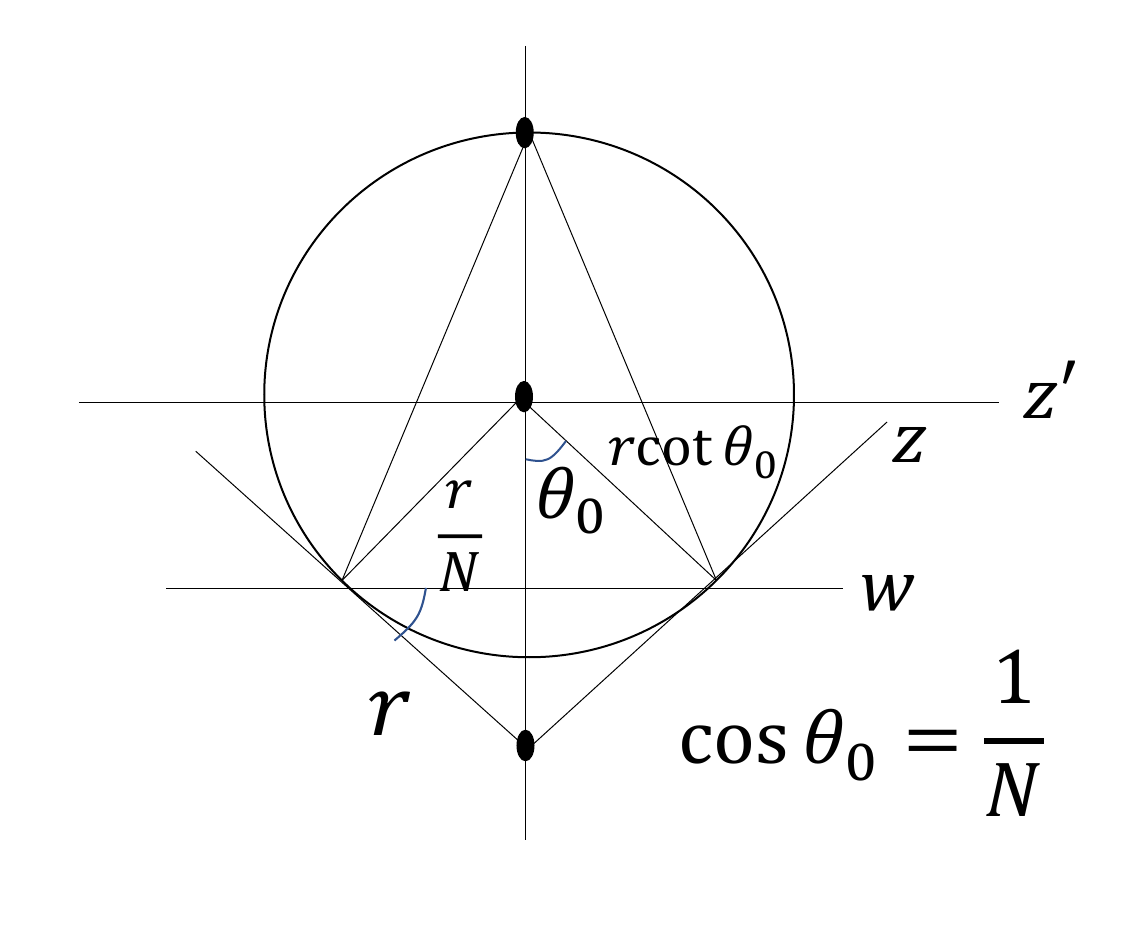}
    \end{minipage}
    \caption{The left figure shows the development of the cone around a singular point of $T^2/\mathbb{Z}_N$ orbifold. The right figure shows the cross section of the cone and the $S^2$ with radius $R =r/\sqrt{N^2-1}$. Here, $z$ and $z'$ denote the coordinates of $T^2/\mathbb{Z}_N$ and $S^2$, respectively, and they are related through the coordinate $w$.}
    \label{fig:ZN}
\end{figure}

The blow-up manifold can be constructed by replacing the cone around the orbifold singular point with the part of $S^2$ such that the total curvature does not change~\cite{Kobayashi:2019fma,Kobayashi:2019gyl,Kobayashi:2022tti}.
Specifically, we first cut the cone with the orbifold singular point whose slant height is $r$.
The left figure in Figure~\ref{fig:ZN} shows the development of the cone and it founds that the radius of the cut surface is $r/N$.
Then, instead of the cone, we smoothly embed $(N-1)/2N$-part of $S^2$ $(0 \leq \theta \leq \theta_0)$ with radius $R=r\cot\theta_0=r/\sqrt{N^2-1}$, where $\cos\theta_0=1/N$.
It can be found from the right figure in Figure~\ref{fig:ZN}, which shows the cross section of the cone and the $S^2$.
Indeed, we can check that the total coverture does not change.
It means that the spin connection on the connected line also does not change.
It can be checked by considering the following coordinate relation.
By introducing $w$ as the coordinate on the cut surface, the coordinate on $T^2/\mathbb{Z}_N$, $z$, and the coordinate on $S^2$, $z'$, are related as
\begin{align}
    z\bigg|_{z=re^{i\varphi/N}} \leftrightarrow w= \frac{N+1}{N}z'\bigg|_{z'=\frac{r}{N+1}e^{i\varphi}},
    \label{eq:coordinaterel}
\end{align}
on the connected line.
In addition, the derivatives on the connected line,
\begin{align}
    \begin{array}{c}
        e^{-i\frac{\varphi}{N}} dz \big|_{z=re^{i\frac{\varphi}{N}}} = d|z| + i r d(\frac{\varphi}{N}), \\
        \frac{N+1}{N} e^{-i\varphi} dz' \big|_{z'=\frac{r}{N+1}e^{i\varphi}} = \frac{N+1}{N} d|z'| + i \frac{r}{N} d\varphi.
    \end{array}
    \label{eq:dzdzprime}
\end{align}
are related as
\begin{align}
    \begin{array}{c}
    e^{-i\frac{\varphi}{N}}dz \bigg|_{z=re^{i\frac{\varphi}{N}}} = \frac{N+1}{N} e^{-i\varphi} dz' \bigg|_{z'=\frac{r}{N+1}e^{i\varphi}} \ 
    \Leftrightarrow \ 
    e^{i\frac{\varphi}{N}} \partial_{z} \bigg|_{z=re^{i\frac{\varphi}{N}}} = \frac{N}{N+1} e^{i\varphi} \partial_{z'} \bigg|_{z'=\frac{r}{N+1}e^{i\varphi}}, \\
    e^{i\frac{\varphi}{N}}d\bar{z} \bigg|_{\bar{z}=re^{-i\frac{\varphi}{N}}} = \frac{N+1}{N} e^{i\varphi} d\bar{z}' \bigg|_{\bar{z}'=\frac{r}{N+1}e^{-i\varphi}} \ 
    \Leftrightarrow \ 
    e^{-i\frac{\varphi}{N}} \partial_{\bar{z}} \bigg|_{\bar{z}=re^{-i\frac{\varphi}{N}}} = \frac{N}{N+1} e^{-i\varphi} \partial_{\bar{z}'} \bigg|_{\bar{z}'=\frac{r}{N+1}e^{-i\varphi}}, 
    \end{array}
    \label{eq:dzdzprimerel}
\end{align}
since it can be found that
\begin{align}
        \frac{N+1}{N} d|z'| = Rd\theta = d|z|, \quad 
        rd\left(\frac{\varphi}{N}\right) = \frac{r}{N} d\varphi,
    \label{derivativerel}
\end{align}
from Figure~\ref{fig:ZN}.
On the blow-up manifold, the $U(1)$ magnetic flux is inserted such that the total magnetic flux does not change from that on the $T^2/\mathbb{Z}_N$ orbifold.
That is, the magnetic flux on the embedded region should be the same as that on the cut out region,
\begin{align}
    \frac{\pi r^2}{N{\rm Im}\tau}M + \frac{N-1}{2N} - \frac{m}{N} + \ell = \frac{N-1}{2N} M' + v'.
    \label{eq:fluxcondition}
\end{align}
In this case, the gauge potential on the connected line also does not change.
In addition, by modifying Eq.~(\ref{eq:fluxcondition}), the following relation is satisfied,
\begin{align}
    \frac{\pi r^2}{N{\rm Im}\tau}M + \frac{(\ell+k)N-(m+k)}{N} = \frac{N-1}{2N}(M'+2k-1) + v'.
    \label{eq:fluxconditionre}
\end{align}
Therefore, by combining Eq.~(\ref{eq:dzdzprimerel}), it is satisfied that,
\begin{align}
    \tilde{{\cal D}}_{(m+k-1,\ell+k-1)}^{(\dagger)} \bigg|_{z=re^{i\frac{\varphi}{N}}} = {\tilde{{\cal D}}}_{(M'+2(k-1))}^{\prime(\dagger)} \bigg|_{z'=\frac{r}{N+1}e^{i\varphi}}.
    \label{eq:calDDprimetilderel}
\end{align}

Now, let us connect wave functions of a spinor smoothly.
Here, we write singular gauge transformed wave functions on magnetized $T^2/\mathbb{Z}_N$ orbifold approximated around $z=0$ in Appendix~\ref{app:T2ZN} and wave functions on magnetized $S^2$ with a vortex at $z'=0$ in Appendix~\ref{app:S2}.
The junction condition of wave functions with level $\forall n$ is given by
\begin{align}
    \tilde{\psi}_{T^2/Z_N^{(m,\ell)},n}^{(\alpha_1,\alpha_{\tau}),M,j}(z) \bigg|_{z=re^{i\varphi/N}} =& {\tilde{\psi}}_{S^2,n}^{\prime M',a'}(z') \bigg|_{z'=\frac{r}{N+1}e^{i\varphi}},
    \label{eq:JCwave} \\
    d\tilde{\psi}_{T^2/Z_N^{(m,\ell)},n}^{(\alpha_1,\alpha_{\tau}),M,j}(z) \bigg|_{z=re^{i\varphi/N}} =& d{\tilde{\psi}}_{S^2,n}^{\prime M',a'}(z') \bigg|_{z'=\frac{r}{N+1}e^{i\varphi}}.
    \label{eq:JCwaveder}
\end{align}
Eq.~(\ref{eq:JCwaveder}) is equivalent to the following two conditions:
\begin{align}
         e^{i\frac{\varphi}{N}} \partial_{z} \tilde{\psi}_{T^2/Z_N^{(m,\ell)},n}^{(\alpha_1,\alpha_{\tau}),M,j}(z) \bigg|_{z=re^{i\varphi/N}} =& \frac{N}{N+1} e^{i\varphi} \partial_{z'}{\tilde{\psi}}_{S^2,n}^{\prime M',a'}(z') \bigg|_{z'=\frac{r}{N+1}e^{i\varphi}} ,
         \label{eq:partialzzprimerel}\\
         e^{-i\frac{\varphi}{N}} \partial_{\bar{z}} \tilde{\psi}_{T^2/Z_N^{(m,\ell)},n}^{(\alpha_1,\alpha_{\tau}),M,j}(z)\bigg|_{\bar{z}=re^{-i\frac{\varphi}{N}}} =& \frac{N}{N+1} e^{-i\varphi} \partial_{\bar{z}'} {\tilde{\psi}}_{S^2,n}^{\prime M',a'}(z') \bigg|_{\bar{z}'=\frac{r}{N+1}e^{-i\varphi}}.
         \label{eq:partialzzprimebarrel}
\end{align}
The following two conditions are equivalent to the above conditions in Eqs.~(\ref{eq:partialzzprimerel}) and (\ref{eq:partialzzprimebarrel}),
\begin{align}
         -i\tilde{{\cal D}}_{(m-1,\ell-1)}^{\dagger} \tilde{\psi}_{T^2/Z_N^{(m,\ell)},n}^{(\alpha_1,\alpha_{\tau}),M,j}(z) \bigg|_{z=re^{i\varphi/N}} =& -i{\tilde{{\cal D}}}_{(M'-2)}^{\prime\dagger}{\tilde{\psi}}_{S^2,n}^{\prime M',a'}(z') \bigg|_{z'=\frac{r}{N+1}e^{i\varphi}},
         \label{eq:DdaggerDdaggerprimerel}\\
         i\tilde{{\cal D}}_{(m,\ell)} \tilde{\psi}_{T^2/Z_N^{(m,\ell)},n}^{(\alpha_1,\alpha_{\tau}),M,j}(z)\bigg|_{z=re^{i\frac{\varphi}{N}}} =& i{\tilde{{\cal D}}^{\prime}}_{(M')} {\tilde{\psi}}_{S^2,n}^{\prime M',a'}(z') \bigg|_{z'=\frac{r}{N+1}e^{i\varphi}}.
         \label{eq:DDprimerel}
\end{align}
Moreover, by applying the equation of motions, they are equivalent to the following two conditions,
\begin{align}
         \tilde{\psi}_{T^2/Z_N^{(m-1,\ell-1)},n+1}^{(\alpha_1,\alpha_{\tau}),M,j}(z) \bigg|_{z=re^{i\varphi/N}} =& {\tilde{\psi}}_{S^2,n+1}^{\prime M'-2,a'-1}(z') \bigg|_{z'=\frac{r}{N+1}e^{i\varphi}},
         \label{eq:levelnp1cond} \\
         {\tilde{{\cal M}}}_n^2(z) \tilde{\psi}_{T^2/Z_N^{(m+1,\ell+1)},n-1}^{(\alpha_1,\alpha_{\tau}),M,j}(z) \bigg|_{z=re^{i\varphi/N}} =& {\tilde{{\cal M}}}_n^{\prime 2}(z') {\tilde{\psi}}_{S^2,n-1}^{\prime M'+2,a'+1}(z') \bigg|_{z'=\frac{r}{N+1}e^{i\varphi}}.
         \label{eq:levelnm1cond}
\end{align}
When Eq.~(\ref{eq:JCwave}) is satisfied for $\forall n$, Eq.~(\ref{eq:levelnp1cond}) is automatically satisfied and Eq.~(\ref{eq:levelnm1cond}) can be rewritten as
\begin{align}
    {\tilde{{\cal M}}}_n^2(z) \bigg|_{z=re^{i\varphi/N}} = {\tilde{{\cal M}}}_n^{\prime 2}(z') \bigg|_{z'=\frac{r}{N+1}e^{i\varphi}}.
    \label{eq:massrel}
\end{align}
To satisfy this condition, it is required that the effective magnetic flux density on the connected line also does not change,~i.e.,
\begin{align}
    \frac{M}{{\rm Im}\tau} = \frac{M'+2k-1}{4\pi R^2}.
    \label{eq:fluxdensitycond}
\end{align}
From Eqs.~(\ref{eq:fluxconditionre}) and (\ref{eq:fluxdensitycond}) with $R=r/\sqrt{N^2-1}$, $M'$ and $v'$ should be satisfied that
\begin{align}
        \frac{N-1}{2N} (M'+2k-1) = \frac{2}{N+1} \frac{\pi r^2}{N{\rm Im}\tau} M, \quad 
        v' = \frac{N-1}{N+1} \frac{\pi r^2}{N{\rm Im}\tau}M + \frac{(\ell+k) N - (m+k)}{N}.
    \label{eq:Mvprime}
\end{align}
Note that the non-zero vortex has to be introduced to connect massive modes smoothly. 
On the other hand, to satisfy Eq.~(\ref{eq:JCwave}), it is required that $a'=\ell$ and the coefficient, ${C'}^{(M',\ell)}$, should be satisfied that
\begin{align}
    C^{\prime(M',\ell)} &= C^{(m,\ell)} \left( \frac{N+1}{2N} \right)^{-\frac{M'-1}{2}} \left( \frac{N-1}{N+1} \right)^{-\frac{\ell-v'}{2}}.
    \label{eq:coefficient}
\end{align}
We summarize the above discussion.
In order for wave functions on magnetized $T^2/\mathbb{Z}_N$ orbifold to connect those on magnetized $S^2$ smoothly, Eqs.~(\ref{eq:JCwave}) and (\ref{eq:massrel}) should be satisfied.
In particular, Eq.~(\ref{eq:massrel}) requires that not only the total magnetic flux as well as the total curvature but also the effective magnetic flux on the connected line do not change under the blow-up procedure.
These conditions are satisfied if we apply wave functions on magnetized $S^2$ with $a'=\ell$, Eqs.~(\ref{eq:Mvprime}), and (\ref{eq:coefficient}).
Indeed, we can check it explicitly by comparing wave functions in Appendix~\ref{app:T2ZN} and \ref{app:S2}.
Therefore, wave functions on the magnetized blow-up manifold are
\begin{align}
    \left\{
    \begin{array}{ll}
       \tilde{\psi}_{T^2/Z_N^{(m,\ell)},\pm,n}^{(\alpha_1,\alpha_{\tau}),M,j}(z)  & (|z| \geq r) \\
       {\tilde{\psi}}_{S^2,\pm,n}^{\prime M',\ell}(z')  & (|z'| \leq \frac{r}{N+1})
    \end{array}
    \right.,
    \label{eq:waveblowup}
\end{align}
with Eqs.~(\ref{eq:Mvprime}), and (\ref{eq:coefficient}).
Although we focus on blowing up the singular point at $z_{\rm f.p.}=0$ in the above analysis, we can similarly blow up the other singular points by replacing the coordinate, $z$, the Scherk-Schwarz phases, $(\alpha_1,\alpha_\tau)$, and the $\mathbb{Z}_N$ charge with $Z$ in Eq.~(\ref{eq:Z}), $(\beta_1,\beta_{\tau})$ in Eq.~(\ref{eq:SSphasebeta}), and $m_{\rm f.p.}$ in Eq.~(\ref{eq:windingnumber}), respectively.


\section{Massive modes of Localized modes}
\label{sec:localizedmode}
So far, we have discussed massive modes of bulk modes.
As shown in section~\ref{sec:T2ZN}, the number of bulk modes of level $n$ with the $\mathbb{Z}_N$ charge, $m$, is the same as that of the bulk zero modes with the $\mathbb{Z}_N$ charge, $m+n$, while the number of localized modes of level $n$ at $z=z_{\rm f.p.}$ is $\ell_{\rm f.p.}+n$.
In this section, we discuss massive modes of localized modes.
In particular, we focus on the localized modes at $z_{\rm f.p.}=0$.
The following discussion can be applied to the other fixed points by replacement appropriately.

To discuss it, we review of bulk modes.
The bulk modes of level $n$ on the magnetized $T^2/\mathbb{Z}_N$, $\tilde{\psi}_{T^2/Z_N^{(m,\ell)},n}^{(\alpha_1,\alpha_{\tau}),M,j}(z)$, constructed by the zero mode wave functions, $\tilde{\psi}_{T^2/Z_N^{(m+n,\ell+n)},0}^{(\alpha_1,\alpha_{\tau}),M,j}(z)$, whose first-order approximation terms are proportional to $z^{(\ell+n)N}$, can be smoothly connected to the level $n$ modes on the magnetized $S^2$ with $a'=\ell$, Eqs.~(\ref{eq:Mvprime}), and (\ref{eq:coefficient}), ${\tilde{\psi}}_{S^2,n}^{\prime\,M',\ell}(z')$, constructed by the zero mode wave functions, ${\tilde{\psi}}_{S^2,0}^{\prime\,M'+2n,\ell+n}(z')$, which are proportional to ${z'}^{\ell+n}$.

As shown in section~\ref{sec:S2}, the level $n$ modes on the magnetized $S^2$ constructed by the zero mode wave functions, ${\tilde{\psi}}_{S^2,0}^{\prime\,M'+2n,a'+n}(z')$, which are proportional to ${z}^{\prime\,a'+n}$, with $-n \leq a' <\ell$ and $\ell < a' < \ell+n$ are independent of ${\tilde{\psi}}_{S^2,n}^{\prime\,M',\ell}(z')$.
 
First, let us consider $\ell \leq a' < \ell+n$ case.
Although the level $n$ modes on the magnetized $S^2$ constructed by the zero mode wave functions, ${\tilde{\psi}}_{S^2,0}^{\prime\,M'+2n,a'+n}(z')$, which are proportional to ${z'}^{a'+n}$,
behave as independent modes in the blow-up region, they connect to the same bulk modes of level $n$ on the magnetized $T^2/\mathbb{Z}_N$, $\tilde{\psi}_{T^2/Z_N^{(m,\ell)},n}^{(\alpha_1,\alpha_{\tau}),M,j}(z)$, constructed by the zero mode wave functions, $\tilde{\psi}_{T^2/Z_N^{(m+n,\ell+n)},0}^{(\alpha_1,\alpha_{\tau}),M,j}(z)$, whose $k$th-order approximation terms are proportional to $z^{(\ell+n+k-1)N}$ with $1 \leq k \leq n$ in the bulk region.
Hence, these modes are not independent on the magnetized $T^2/\mathbb{Z}_N$ orbifold.

Next, let us consider $-n \leq a' <\ell$ case.
The level $n$ modes on the magnetized $S^2$ constructed by the zero mode wave functions, ${\tilde{\psi}}_{S^2,0}^{\prime\,M'+2n,a'+n}(z')$, which are proportional to ${z'}^{a'+n}$, can connect to the level $n$ modes on the magnetized $T^2/\mathbb{Z}_N$ orbifold constructed by the zero mode wave functions whose first-order approximation terms are proportional to $z^{(a'+n)N}$.
Indeed, since operating $(J^{\prime -}_{n})^{k} \ (k \leq \ell+n)$, defined in Appendix~\ref{app:AM}, to ${\tilde{\psi}}_{S^2,0}^{\prime\,M'+2n,\ell+n}(z')$ only changes $z^{\prime\ell+n}$ to $z^{\prime\ell+n-k}$ times the overall factor, the level $n$ modes constructed by $(J^{\prime -}_{n})^{k}{\tilde{\psi}}_{S^2,0}^{\prime\,M'+2n,a'+n}(z')$ have same mass eigenvalue of those constructed by ${\tilde{\psi}}_{S^2,0}^{\prime\,M'+2n,a'+n}(z')$.
Similarly, operating $(J^{-}_{n})^{k}$, defined in Appendix~\ref{app:AM}, to $\tilde{\psi}_{T^2/Z_N^{(m+n,\ell+n)},0}^{(\alpha_1,\alpha_{\tau}),M,j}(z)$ only changes $z^{(\ell+n)N}$ to $z^{(\ell+n-k)N}$ times the overall factor, and then the level $n$ modes constructed by $(J^{-}_{n})^{k}\tilde{\psi}_{T^2/Z_N^{(m+n,\ell+n)},0}^{(\alpha_1,\alpha_{\tau}),M,j}(z)$ have same mass eigenvalue of those constructed by ${\tilde{\psi}}_{S^2,0}^{\prime\,M'+2n,a'+n}(z')$.
Note that, as in Ref.~\cite{Kobayashi:2022tti}, $(J^{-}_{n})^{k}\tilde{\psi}_{T^2/Z_N^{(m+n,\ell+n)},0}^{(\alpha_1,\alpha_{\tau}),M,j}(z)$ as well as the level $n$ modes constructed by them become localized modes.
In addition, by including the overall factor, we can check that
\begin{align}
    (J^{-}_{n})^{k}\tilde{\psi}_{T^2/Z_N^{(m+n,\ell+n)},0}^{(\alpha_1,\alpha_{\tau}),M,j}(z) \bigg|_{\bar{z}=re^{-i\frac{\varphi}{N}}} = (J^{\prime -}_{n})^{k}{\tilde{\psi}}_{S^2,0}^{\prime\,M'+2n,a'+n}(z') \bigg|_{\bar{z}'=\frac{r}{N+1}e^{-i\varphi}}.
    \label{eq:J-psi}
\end{align}
Therefore, the level $n$ modes constructed by them can be also smoothly connected.

In summary, there are $(\ell+n)$-numbers of localized modes of level $n$ constructed by the zero mode wave functions, $(J^{-}_{n})^{k}\tilde{\psi}_{T^2/Z_N^{(m+n,\ell+n)},0}^{(\alpha_1,\alpha_{\tau}),M,j}(z) \ (1\leq k\leq \ell+n)$, instead of $\tilde{\psi}_{T^2/Z_N^{(m+n,\ell+n)},0}^{(\alpha_1,\alpha_{\tau}),M,j}(z)$, and they can smoothly connect to the level $n$ modes constructed by the zero mode wave functions, $(J^{\prime -}_{n})^{k}{\tilde{\psi}}_{S^2,0}^{\prime\,M'+2n,a'+n}(z')$, instead of ${\tilde{\psi}}_{S^2,0}^{\prime\,M'+2n,a'+n}(z')$. 
This behavior is similar to $SU(2)$ Kac-Moody algebra \cite{Goddard:1986bp}, but there is a difference.
The representations are restricted universally for all modes $n$ by the level in the Kac-Moody algebra. 
However, the angular momentum of localized modes are restricted by $\ell+n$, which depends on the mode number $n$.

Magnetized D-brane models are T-dual to intersecting D-brane models.
Furthermore, conformal field theoretical aspects of 
intersecting D-brane models \cite{Cvetic:2003ch,Abel:2003vv} are similar to those of heterotic orbifold models \cite{Hamidi:1986vh,Dixon:1986qv}.
Heterotic orbifold models and intersecting D-brane models also have towers of localized massive modes at orbifold fixed points and intersecting points. 
It is interesting to compare localized massive modes in our models with those in intersecting D-brane models and heterotic orbifold models.
That is beyond our scope.


\section{Conclusion}
\label{sec:conclusion}

In this paper, we have studied massive modes on the magnetized blow-up manifold of $T^2/\mathbb{Z}_N$. 
The blow-up manifold can be constructed by replacing orbifold singular points with a part of $S^2$ such that the total curvature does not change. 
As discussed in Refs.~\cite{Kobayashi:2019fma,Kobayashi:2022tti}, smoothly connecting the zero modes on magnetized $T^2/\mathbb{Z}_N$ orbifold with those on the magnetized $S^2$ on the connected line requires that the total magnetic flux does not change. 
In section~\ref{sec:blowupmanifold}, we have found that it requires that not only the total magnetic flux as well as the total curvature but also the effective magnetic flux on the connected line do not change under the blow-up procedure so that mass eigenvalues remain unchanged on the connected line. 
To satisfy these conditions, we have introduced a magnetized $S^2$ with a vortex. 
Specifically, we have constructed the massive modes on the blow-up manifold by applying appropriate magnetic flux, vortex, and coefficients as shown in Eqs.~(\ref{eq:Mvprime}) and (\ref{eq:coefficient}). 
The construction of massive modes on magnetized $T^2/\mathbb{Z}_N$ orbifold after the singular gauge transformation and that on magnetized $S^2$ with a vortex have been discussed in sections~\ref{sec:T2ZN} and \ref{sec:S2}, respectively, while the explicit wave functions of massive modes on magnetized $T^2/\mathbb{Z}_N$ orbifold and magnetized $S^2$ are shown in Appendices~\ref{app:T2ZN} and \ref{app:S2}, respectively.

In addition, in section~\ref{sec:localizedmode}, we have found that the number of degenerated states increases by the number of orbifold singular points with each increment of the mass label. These states correspond to modes localized at the singular points. 
We demonstrated that they can be constructed by replacing the bulk zero mode with the zero mode obtained by operating the lowering operator of the angular momentum to the bulk zero mode, where the ladder operators of the angular momentum on the magnetized $T^2/\mathbb{Z}_N$ can be defined from those on the magnetized $S^2$ through the blow-up procedure as shown in Appendix~\ref{app:AM}.

As for future work, we would like to study phenomenological implications of this construction, including loop-level effects on the blow-up manifold such as threshold corrections due to massive modes on gauge and Yukawa couplings. 
It is interesting to compare the towers of localized massive modes in our models with those in intersecting D-brane models and heterotic orbifold models. 
Furthermore, we would also like to extend this blow-up procedure to other higher-dimensional orbifolds.

\section*{Acknowledgments}
This work was supported in part by JSPS KAKENHI Grant Numbers JP23K03375 (T. K.),  JP25H01539 (H.O.) and JP26K07087 (H.O.).

\appendix


\section{Angular momentum operator on magnetized $T^2/\mathbb{Z}_N$}
\label{app:AM}

Here, we define the angular momentum operators on magnetized $T^2/\mathbb{Z}_N$ orbifold from those on magnetized $S^2$ through the blow-up procedure.

We define the angular momentum operators on magnetized $T^2/\mathbb{Z}_N$ orbifold as
\begin{align}
    J_k^+
    =& \frac{1}{L^N} \left( \frac{z^{2N}}{Nz^{N-1}} \tilde{D}_{z}^{(m+k,\ell+k)} + \frac{L^{2N}}{N\bar{z}^{N-1}} \tilde{D}_{\bar{z}}^{(m+k,\ell+k)} - \frac{L^{2N}}{N\bar{z}^{N-1}} \left( \frac{\pi M}{{\rm Im}\tau}z + \frac{(\ell+k)N-(m+k))}{\bar{z}} \right) \right),
    \label{eq:J+} \\
    J_k^-
    =& - \frac{1}{L^N} \left( \frac{\bar{z}^{2N}}{N\bar{z}^{N-1}} \tilde{D}_{\bar{z}}^{(m+k,\ell+k)} + \frac{L^{2N}}{N z^{N-1}} \tilde{D}_{z}^{(m+k,\ell+k)} + \frac{L^{2N}}{N z^{N-1}} \left( \frac{\pi M}{{\rm Im}\tau}\bar{z} + \frac{(\ell+k)N-(m+k))}{z} \right) \right),
    \label{eq:J-} \\
    J_k^3
    =& \frac{z}{N} \partial_{z} - \frac{\bar{z}}{N} \partial_{\bar{z}} - \frac{2N}{N^2-1} \frac{\pi r^2}{N{\rm Im}\tau} M,
    \label{eq:J3}
\end{align}
with $L^N=r^N \sqrt{\frac{N+1}{N-1}}$
such that they correspond to those on magnetized $S^2$ in Eqs.~(\ref{eq:Jprime+})-(\ref{eq:Jprime3}) on the connected line,~i.e.,
\begin{align}
    J^{a}_{k} \bigg|_{z=re^{i\frac{\varphi}{N}}} = J^{\prime a}_k \bigg|_{z'=\frac{r}{N+1}e^{i\varphi}} \quad (a=+, -,3).
    \label{eq:Jcondition}
\end{align}
Here, we take into account the following relations,
\begin{align}
    \begin{array}{c}
        z^N \partial_{z^N} \bigg|_{z=re^{i\frac{\varphi}{N}}} = \frac{1}{N} z \partial_{z} \bigg|_{z=re^{i\frac{\varphi}{N}}} = z' \partial_{z'} \bigg|_{z'=\frac{r}{N+1}e^{i\varphi}}, \\
        \bar{z}^N \partial_{\bar{z}^N} \bigg|_{\bar{z}=re^{-i\frac{\varphi}{N}}} = \frac{1}{N} \bar{z} \partial_{\bar{z}} \bigg|_{\bar{z}=re^{-i\frac{\varphi}{N}}} = \bar{z}' \partial_{\bar{z}'} \bigg|_{\bar{z}'=\frac{r}{N+1}e^{-i\varphi}}.
    \end{array}
    \label{eq:zpartialz}
\end{align}
Indeed, their algebraic relations on the connected line are consistent with Eqs.~(\ref{eq:primeJ+J-})-(\ref{eq:primeJ3J-}).


\section{Explicit wave functions on magnetized $T^2/\mathbb{Z}_N$ orbifold}
\label{app:T2ZN}

Here, we show explicit wave functions on magnetized $T^2/\mathbb{Z}_N$ orbifold after the singular gauge transformation approximated around $z=0$.

First, the wave function of level $0$ with positive chirality is given by
\begin{align}
    \tilde{\psi}_{T^2/Z_N^{(m,\ell)},+,0}^{(\alpha_1,\alpha_{\tau}),M,j}(z) &= \tilde{\psi}_{T^2/Z_N^{(m,\ell)},0}^{(\alpha_1,\alpha_{\tau}),M,j}(z) \notag \\
    &\simeq C^{(m,\ell)} e^{-\frac{\pi M}{2{\rm Im}\tau}(|z|^2-r^2)} \left( \frac{|z|}{r} \right)^{-(\ell N-m)} \left(\frac{z}{r} \right)^{\ell N},
    \label{eq:+0T2ZN} \\
    \tilde{\psi}_{T^2/Z_N^{(m,\ell)},+,0}^{(\alpha_1,\alpha_{\tau}),M,j}(z=re^{i\frac{\varphi}{N}})
    &\simeq C^{(m,\ell)} e^{i\ell\varphi}.
    \label{eq:+0T2ZNboundary}
\end{align}
Similarly, the wave function of level $1$ with negative chirality is given by
\begin{align}
    \tilde{\psi}_{T^2/Z_N^{(m,\ell)},-,1}^{(\alpha_1,\alpha_{\tau}),M,j}(z)
    &= \tilde{\psi}_{T^2/Z_N^{(m+1,\ell+1)},0}^{(\alpha_1,\alpha_{\tau}),M,j}(z) \notag \\
    &\simeq C^{(m+1,\ell+1)} e^{-\frac{\pi M}{2{\rm Im}\tau}(|z|^2-r^2)} \left( \frac{|z|}{r} \right)^{-((\ell+1) N-(m+1))} \left(\frac{z}{r} \right)^{(\ell +1)N},
    \label{eq:-1T2ZN} \\
    \tilde{\psi}_{T^2/Z_N^{(m,\ell)},-,1}^{(\alpha_1,\alpha_{\tau}),M,j}(z=re^{i\frac{\varphi}{N}})
    &\simeq C^{(m+1,\ell+1)} e^{i(\ell+1)\varphi},
    \label{eq:-1T2ZNboundary}
\end{align}
where we note that it should be changed from $(m,\ell)$ to $(m+1,\ell+1)$.
Next, the wave function of level $1$ with positive chirality is given by
\begin{align}
    \tilde{\psi}_{T^2/Z_N^{(m,\ell)},+,1}^{(\alpha_1,\alpha_{\tau}),M,j}(z)
    =& \tilde{\psi}_{T^2/Z_N^{(m,\ell)},1}^{(\alpha_1,\alpha_{\tau}),M,j}(z) \notag \\
    -i \tilde{{\cal D}}^{\dagger}_{(m,\ell)} \tilde{\psi}_{T^2/Z_N^{(m,\ell)},-,1}^{(\alpha_1,\alpha_{\tau}),M,j}(z) 
    =& -i \tilde{{\cal D}}^{\dagger}_{(m,\ell)} \tilde{\psi}_{T^2/Z_N^{(m+1,\ell+1)},0}^{(\alpha_1,\alpha_{\tau}),M,j}(z) \notag \\
    \simeq& i \left( (\ell+1) - \left( \frac{\pi M}{N{\rm Im}\tau}|z|^2 + \frac{(\ell +1)N -(m+1)}{N} \right) \right) \notag \\
    &\times \frac{2N}{r} C^{(m+1,\ell+1)} e^{-\frac{\pi M}{2{\rm Im}\tau}(|z|^2-r^2)} \left( \frac{|z|}{r} \right)^{-(\ell N-m)} \left(\frac{z}{r} \right)^{\ell N},
    \label{eq:+1T2ZN} \\
    \tilde{\psi}_{T^2/Z_N^{(m,\ell)},+,1}^{(\alpha_1,\alpha_{\tau}),M,j}(z=re^{i\frac{\varphi}{N}})
    \simeq&  i \left( (\ell+1) - \left( \frac{\pi r^2}{N{\rm Im}\tau}M + \frac{(\ell +1)N -(m+1)}{N} \right) \right) \frac{2N}{r} C^{(m+1,\ell+1)} e^{i\ell\varphi}.
    \label{eq:+1T2ZNboundary}
\end{align}
Similarly, the wave function of level $2$ with negative chirality is given by
\begin{align}
    \tilde{\psi}_{T^2/Z_N^{(m,\ell)},-,2}^{(\alpha_1,\alpha_{\tau}),M,j}(z)
    =& \tilde{\psi}_{T^2/Z_N^{(m+1,\ell+1)},1}^{(\alpha_1,\alpha_{\tau}),M,j}(z) = -i \tilde{{\cal D}}^{\dagger}_{(m+1,\ell+1)} \tilde{\psi}_{T^2/Z_N^{(m+2,\ell+2)},0}^{(\alpha_1,\alpha_{\tau}),M,j}(z) \notag \\
    \simeq& \left( (\ell+2) - \left( \frac{\pi M}{N{\rm Im}\tau}|z|^2 + \frac{(\ell +2)N -(m+2)}{N} \right) \right) \notag \\
    &\times i \frac{2N}{r} C^{(m+2,\ell+2)} e^{-\frac{\pi M}{2{\rm Im}\tau}(|z|^2-r^2)} \left( \frac{|z|}{r} \right)^{-((\ell+1) N-(m+1))} \left(\frac{z}{r} \right)^{(\ell +1)N},
    \label{eq:-2T2ZN} \\
    \tilde{\psi}_{T^2/Z_N^{(m,\ell)},-,2}^{(\alpha_1,\alpha_{\tau}),M,j}(z=re^{i\frac{\varphi}{N}})
    \simeq&  \left( (\ell+2) - \left( \frac{\pi r^2}{N{\rm Im}\tau}M + \frac{(\ell +2)N -(m+2)}{N} \right) \right) i \frac{2N}{r} C^{(m+2,\ell+2)} e^{i(\ell+1)\varphi}.
    \label{eq:-2T2ZNboundary}
\end{align}
The wave function of level $2$ with positive chirality is given by
\begin{align}
    \tilde{\psi}_{T^2/Z_N^{(m,\ell)},+,2}^{(\alpha_1,\alpha_{\tau}),M,j}(z) =& \tilde{\psi}_{T^2/Z_N^{(m,\ell)},2}^{(\alpha_1,\alpha_{\tau}),M,j}(z) \notag \\
    -i \tilde{{\cal D}}^{\dagger}_{(m,\ell)} \tilde{\psi}_{T^2/Z_N^{(m,\ell)},-,2}^{(\alpha_1,\alpha_{\tau}),M,j}(z) =& (-i)^2 \tilde{{\cal D}}^{\dagger}_{(m,\ell)} \tilde{{\cal D}}^{\dagger}_{(m+1,\ell+1)} \tilde{\psi}_{T^2/Z_N^{(m+2,\ell+2)},0}^{(\alpha_1,\alpha_{\tau}),M,j}(z) \notag \\
    \simeq& \left\{ \prod_{k=1}^{2} \left( (\ell+k) - \left( \frac{\pi M}{N{\rm Im}\tau}|z|^2 + \frac{(\ell +k)N -(m+k)}{N} \right) \right) - \frac{1}{N} \frac{\pi M}{N{\rm Im}\tau}|z|^2 \right\} \notag \\
    &\times \left( i \frac{2N}{r} \right)^2 C^{(m+2,\ell+2)} e^{-\frac{\pi M}{2{\rm Im}\tau}(|z|^2-r^2)} \left( \frac{|z|}{r} \right)^{-(\ell N-m)} \left(\frac{z}{r} \right)^{\ell N},
    \label{eq:+2T2ZN}
\end{align}
\begin{align}
    &\tilde{\psi}_{T^2/Z_N^{(m,\ell)},+,2}^{(\alpha_1,\alpha_{\tau}),M,j}(z=re^{i\frac{\varphi}{N}}) \notag \\
    \simeq& \left\{ \prod_{k=1}^{2} \left( (\ell+k) - \left( \frac{\pi r^2}{N{\rm Im}\tau}M + \frac{(\ell +k)N -(m+k)}{N} \right) \right) - \frac{1}{N} \frac{\pi r^2}{N{\rm Im}\tau}M \right\} \notag \\
    &\times \left( i \frac{2N}{r} \right)^2 C^{(m+2,\ell+2)} e^{i\ell\varphi}.
    \label{eq:+2T2ZNboundary}
\end{align}
Similarly, the wave function of level $3$ with negative chirality is given by
\begin{align}
    &\tilde{\psi}_{T^2/Z_N^{(m,\ell)},-,3}^{(\alpha_1,\alpha_{\tau}),M,j}(z)
    = \tilde{\psi}_{T^2/Z_N^{(m+1,\ell+1)},2}^{(\alpha_1,\alpha_{\tau}),M,j}(z) =  (-i)^3 \prod_{k=2}^{3} \tilde{{\cal D}}^{\dagger}_{(m+k-1,\ell+k-1)} \tilde{\psi}_{T^2/Z_N^{(m+3,\ell+3)},0}^{(\alpha_1,\alpha_{\tau}),M,j}(z) \notag \\
    \simeq& \left\{ \prod_{k=2}^{3} \left( (\ell+k) - \left( \frac{\pi M}{N{\rm Im}\tau}|z|^2 + \frac{(\ell +k)N -(m+k)}{N} \right) \right) - \frac{1}{N} \frac{\pi M}{N{\rm Im}\tau}|z|^2 \right\} \notag \\
    &\times \left( i \frac{2N}{r} \right)^2 C^{(m+3,\ell+3)} e^{-\frac{\pi M}{2{\rm Im}\tau}(|z|^2-r^2)} \left( \frac{|z|}{r} \right)^{-((\ell+1) N-(m+1))} \left(\frac{z}{r} \right)^{(\ell +1)N},
    \label{eq:-3T2ZN} \\
    &\tilde{\psi}_{T^2/Z_N^{(m,\ell)},-,3}^{(\alpha_1,\alpha_{\tau}),M,j}(z=re^{i\frac{\varphi}{N}}) \notag \\
    \simeq& \left\{ \prod_{k=2}^{3} \left( (\ell+k) - \left( \frac{\pi r^2}{N{\rm Im}\tau}M + \frac{(\ell +k)N -(m+k)}{N} \right) \right) - \frac{1}{N} \frac{\pi r^2}{N{\rm Im}\tau}M \right\} \notag \\
    &\times \left( i \frac{2N}{r} \right)^2 C^{(m+3,\ell+3)} e^{i(\ell +1)\varphi}.
    \label{eq:-3T2ZNboundary}
\end{align}
The wave function of level $3$ with positive chirality is given by
\begin{align}
    &\tilde{\psi}_{T^2/Z_N^{(m,\ell)},+,3}^{(\alpha_1,\alpha_{\tau}),M,j}(z) = \tilde{\psi}_{T^2/Z_N^{(m,\ell)},3}^{(\alpha_1,\alpha_{\tau}),M,j}(z) \notag \\
    &-i \tilde{{\cal D}}^{\dagger}_{(m,\ell)} \tilde{\psi}_{T^2/Z_N^{(m,\ell)},-,3}^{(\alpha_1,\alpha_{\tau}),M,j}(z) = (-i)^3 \prod_{k=1}^{3} \tilde{{\cal D}}^{\dagger}_{(m+k-1,\ell+k-1)} \tilde{\psi}_{T^2/Z_N^{(m+3,\ell+3)},0}^{(\alpha_1,\alpha_{\tau}),M,j}(z) \notag \\
    \simeq& \biggl\{
    \prod_{k=1}^{3} \left( (\ell+k) - \left( \frac{\pi M}{N{\rm Im}\tau}|z|^2 + \frac{(\ell +k)N -(m+k)}{N} \right) \right) \notag \\
    &- \frac{1}{N} \frac{\pi M}{N{\rm Im}\tau}|z|^2 \sum_{k=1}^{3} \left( (\ell+k) - \left( \frac{\pi M}{N{\rm Im}\tau}|z|^2 + \frac{(\ell +k)N -(m+k)}{N} \right) \right) - \frac{1}{N^2} \frac{\pi M}{N{\rm Im}\tau}|z|^2
    \biggr\} \notag \\
    &\times \left( i \frac{2N}{r} \right)^3 C^{(m+3,\ell+3)} e^{-\frac{\pi M}{2{\rm Im}\tau}(|z|^2-r^2)} \left( \frac{|z|}{r} \right)^{-(\ell N-m)} \left(\frac{z}{r} \right)^{\ell N},
    \label{eq:+3T2ZN} \\
    &\tilde{\psi}_{T^2/Z_N^{(m,\ell)},+,3}^{(\alpha_1,\alpha_{\tau}),M,j}(z=re^{i\frac{\varphi}{N}}) \notag \\
    \simeq& \biggl\{
    \prod_{k=1}^{3} \left( (\ell+k) - \left( \frac{\pi r^2}{N{\rm Im}\tau}M + \frac{(\ell +k)N -(m+k)}{N} \right) \right) \notag \\
    &- \frac{1}{N} \frac{\pi r^2}{N{\rm Im}\tau}M \sum_{k=1}^{3} \left( (\ell+k) - \left( \frac{\pi r^2}{N{\rm Im}\tau}M + \frac{(\ell +k)N -(m+k)}{N} \right) \right) - \frac{1}{N^2} \frac{\pi r^2}{N{\rm Im}\tau}M
    \biggr\} \notag \\
    &\times \left( i \frac{2N}{r} \right)^3 C^{(m+3,\ell+3)} e^{i\ell\varphi}.
    \label{eq:+3T2ZNboundary}
\end{align}
In this way, wave functions of level $n$ with the positive and negative chiralities are given in Eqs.~(\ref{eq:psiT2ZN+ntilde}) and (\ref{eq:psiT2ZN-ntilde}), respectively.


\section{Explicit wave functions on magnetized $S^2$}
\label{app:S2}

Here, we show explicit wave functions on magnetized $S^2$ with vortex, $v'$, at $z'=0$.

First, the wave function of level $0$ with positive chirality is given by
\begin{align}
    {\tilde{\psi}}_{S^2,+,0}^{\prime\,M',a'}(z') &= {\tilde{\psi}}_{S^2,0}^{\prime\,M',a'}(z') \notag \\
    &= C^{\prime(M',a')} \left( \frac{R^2}{R^2+|z'|^2} \right)^{\frac{M'-1}{2}} \left( \frac{|z'|}{R} \right)^{-v'} \left( \frac{z'}{R} \right)^{a'},
    \label{eq:+0S2}
\end{align}
\begin{align}
    {\tilde{\psi}}_{S^2,+,0}^{\prime\,M',a'}(z'=\frac{r}{N+1}e^{i\varphi}, R=\frac{r}{\sqrt{N^2-1}})
    &= C^{\prime(M',a')} \left( \frac{N+1}{2N} \right)^{\frac{M'-1}{2}} \left( \frac{N-1}{N+1} \right)^{\frac{a' - v'}{2}} e^{ia'\varphi}.
    \label{eq:+0S2boundary}
\end{align}
Similarly, the wave function of level $1$ with negative chirality is given by
\begin{align}
    {\tilde{\psi}}_{S^2,-,1}^{\prime\,M',a'}(z')
    &= {\tilde{\psi}}_{S^2,0}^{\prime\,M'+2,a'+1}(z') \notag \\
    &=C^{\prime(M'+2,a'+1)} \left( \frac{R^2}{R^2+|z'|^2} \right)^{\frac{M'+1}{2}} \left( \frac{|z'|}{R} \right)^{-v'} \left( \frac{z'}{R} \right)^{a'+1},
    \label{eq:-1S2}
\end{align}
\begin{align}
    {\tilde{\psi}}_{S^2,-,1}^{\prime\,M',a'}(z'=\frac{r}{N+1}e^{i\varphi}, R=\frac{r}{\sqrt{N^2-1}})
    &= C^{\prime(M'+2,a'+1)} \left( \frac{N+1}{2N} \right)^{\frac{M'+1}{2}} \left( \frac{N-1}{N+1} \right)^{\frac{a'+1-v'}{2}} e^{i(a'+1)\varphi},
    \label{eq:-1S2boundary}
\end{align}
where we note that it should be changed from $M'-1$ to $M'+1$.
Next, the wave function of level $1$ with positive chirality is given by
\begin{align}
    {\tilde{\psi}}_{S^2,+,1}^{\prime\,M',a'}(z') =& {\tilde{\psi}}_{S^2,1}^{\prime\,M',a'}(z') \notag \\
    -i {\tilde{{\cal D}}'}_{(M')} {\tilde{\psi}}_{S^2,-,1}^{\prime\,M',a'}(z') =&  -i {\tilde{{\cal D}}'}_{(M')} {\tilde{\psi}}_{S^2,0}^{\prime\,M'+2,a'+1}(z') \notag \\
    =& i \left( (a'+1) - \left( \frac{|z'|^2}{R^2+|z'|^2}(M'+1) + v' \right) \right) \notag \\
    &\times \frac{1}{R} C^{\prime(M'+2,a'+1)} \left( \frac{R^2}{R^2+|z'|^2} \right)^{\frac{M'-1}{2}} \left( \frac{|z'|}{R} \right)^{-v'} \left( \frac{z'}{R} \right)^{a'},
    \label{eq:+1S2}
\end{align}
\begin{align}
    &{\tilde{\psi}}_{S^2,+,1}^{\prime\,M',a'}(z'=\frac{r}{N+1}e^{i\varphi}, R=\frac{r}{\sqrt{N^2-1}}) \notag \\
    =& i \left( (a'+1) - \left( \frac{N-1}{2N}(M'+1) +v' \right) \right) \frac{\sqrt{N^2-1}}{r} C^{\prime(M'+2,a'+1)} \left( \frac{N+1}{2N} \right)^{\frac{M'-1}{2}} \left( \frac{N-1}{N+1} \right)^{\frac{a'-v'}{2}} e^{ia'\varphi}.
    \label{eq:+1S2boundary}
\end{align}
Similarly, the wave function of level $2$ with negative chirality is given by
\begin{align}
    &{\tilde{\psi}}_{S^2,-,2}^{\prime\,M',a'}(z')
    = {\tilde{\psi}}_{S^2,1}^{\prime\,M'+2,a'+1}(z') = -i {\tilde{{\cal D}}'}_{(M'+2)} {\tilde{\psi}}_{S^2,0}^{\prime\,M'+4,a'+2}(z') \notag \\
    =& i \left( (a'+2) - \left( \frac{|z'|^2}{R^2+|z'|^2}(M'+3) + v' \right) \right) \frac{1}{R} C^{\prime(M'+4,a'+2)} \left( \frac{R^2}{R^2+|z'|^2} \right)^{\frac{M'+1}{2}} \left( \frac{|z'|}{R} \right)^{-v'} \left( \frac{z'}{R} \right)^{a'+1},
    \label{eq:-2S2}
\end{align}
\begin{align}
    &{\tilde{\psi}}_{S^2,-,2}^{\prime\,M',a'}(z'=\frac{r}{N+1}e^{i\varphi}, R=\frac{r}{\sqrt{N^2-1}}) \notag \\
    =& i \left( (a'+2) - \left( \frac{N-1}{2N}(M'+3) + v' \right) \right) \notag \\
    &\times \frac{\sqrt{N^2-1}}{r} C^{\prime(M'+4,a'+2)} \left( \frac{N+1}{2N} \right)^{\frac{M'+1}{2}} \left( \frac{N-1}{N+1} \right)^{\frac{a'+1-v'}{2}} e^{i(a'+1)\varphi}.
    \label{eq:-2S2boundary}
\end{align}
The wave function of level $2$ with positive chirality is given by
\begin{align}
    &{\tilde{\psi}}_{S^2,+,2}^{\prime\,M',a'}(z') ={\tilde{\psi}}_{S^2,2}^{\prime\,M',a'}(z') \notag \\
    &-i {\tilde{{\cal D}}'}_{(M')} {\tilde{\psi}}_{S^2,-,2}^{\prime\,M',a'}(z') = (-i)^2 {\tilde{{\cal D}}'}_{(M')} {\tilde{{\cal D}}'}_{(M'+2)} {\tilde{\psi}}_{S^2,0}^{\prime\,M'+4,a'+2}(z') \notag \\
    =& \left\{ \prod_{k=1}^{2} \left( (a'+k) - \left( \frac{|z'|^2}{R^2+|z'|^2}(M'+2k-1) + v' \right) \right) - \frac{R^2}{R^2+|z'|^2}\frac{|z'|^2}{R^2+|z'|^2}(M'+3) \right\} \notag \\
    &\times \left( i \frac{1}{R} \right)^2 C^{\prime(M'+4,a'+2)} \left( \frac{R^2}{R^2+|z'|^2} \right)^{\frac{M'-1}{2}} \left( \frac{|z'|}{R} \right)^{-v'} \left( \frac{z'}{R} \right)^{a'},
    \label{eq:+2S2}
\end{align}
\begin{align}
    &{\tilde{\psi}}_{S^2,+,2}^{\prime\,M',a'}(z'=\frac{r}{N+1}e^{i\varphi}, R=\frac{r}{\sqrt{N^2-1}}) \notag \\
    =& \left\{ \prod_{k=1}^{2} \left( (a'+k) - \left( \frac{N-1}{2N}(M'+2k-1) + v' \right) \right) - \frac{N+1}{2N}\frac{N-1}{2N}(M'+3) \right\} \notag \\
    &\times \left( i \frac{1}{R} \right)^2 C^{\prime(M'+4,a'+2)} \left( \frac{N+1}{2N} \right)^{\frac{M'-1}{2}} \left( \frac{N-1}{N+1} \right)^{\frac{a'-v'}{2}} e^{ia'\varphi}.
    \label{eq:+2S2boundary}
\end{align}
Similarly, the wave function of level $3$ with negative chirality is given by
\begin{align}
    &{\tilde{\psi}}_{S^2,-,3}^{\prime\,M',a'}(z')
    = {\tilde{\psi}}_{S^2,2}^{\prime\,M'+2,a'+1}(z') = (-i)^2 \prod_{k=2}^{3} {\tilde{{\cal D}}'}_{(M'+2(k-1))} {\tilde{\psi}}_{S^2,0}^{\prime\,M'+6,a'+3}(z') \notag \\
    =& \left\{ \prod_{k=2}^{3} \left( (a'+k) - \left( \frac{|z'|^2}{R^2+|z'|^2}(M'+2k-1) + v' \right) \right) - \frac{R^2}{R^2+|z'|^2}\frac{|z'|^2}{R^2+|z'|^2}(M'+5) \right\} \notag \\
    &\times \left( i \frac{1}{R} \right)^2 C^{\prime(M'+6,a'+3)} \left( \frac{R^2}{R^2+|z'|^2} \right)^{\frac{M'+1}{2}} \left( \frac{|z'|}{R} \right)^{-v'} \left( \frac{z'}{R} \right)^{a'+1},
    \label{eq:-3S2}
\end{align}
\begin{align}
    &{\tilde{\psi}}_{S^2,-,3}^{\prime\,M',a'}(z'=\frac{r}{N+1}e^{i\varphi}, R=\frac{r}{\sqrt{N^2-1}}) \notag \\
    =& \left\{ \prod_{k=2}^{3} \left( (a'+k) - \left( \frac{N-1}{2N}(M'+2k-1) + v' \right) \right) - \frac{N+1}{2N}\frac{N-1}{2N}(M'+5) \right\} \notag \\
    &\times \left( i \frac{1}{R} \right)^2 C^{\prime(M'+6,a'+4)} \left( \frac{N+1}{2N} \right)^{\frac{M'+1}{2}} \left( \frac{N-1}{N+1} \right)^{\frac{a'+1-v'}{2}} e^{i(a'+1)\varphi}.
    \label{eq:-3S2boundary}
\end{align}
The wave function of level $3$ with positive chirality is given by
\begin{align}
    {\tilde{\psi}}_{S^2,+,3}^{\prime\,M',a'}(z') =& {\tilde{\psi}}_{S^2,3}^{\prime\,M',a'}(z') \notag \\
    -i {\tilde{{\cal D}}'}_{(M')} {\tilde{\psi}}_{S^2,-,3}^{\prime\,M',a'}(z') =& (-i)^3 \prod_{k=1}^{3} {\tilde{{\cal D}}'}_{(M'+2(k-1))} {\tilde{\psi}}_{S^2,0}^{\prime\,M'+6,a'+3}(z') \notag \\
    =& \biggl\{ \prod_{k=1}^{3} \left( (a'+k) - \left( \frac{|z'|^2}{R^2+|z'|^2}(M'+2k-1) + v' \right) \right) \notag \\
    &- \frac{R^2}{R^2+|z'|^2}\frac{|z'|^2}{R^2+|z'|^2}(M'+5) \left( (a'+1) - \left( \frac{|z'|^2}{R^2+|z'|^2}(M'+1) + v' \right) \right) \notag \\
    &- \frac{R^2}{R^2+|z'|^2}\frac{|z'|^2}{R^2+|z'|^2}(M'+5) \left( (a'+2) - \left( \frac{|z'|^2}{R^2+|z'|^2}(M'+3) + v' \right) \right) \notag \\
    &- \frac{R^2}{R^2+|z'|^2}\frac{|z'|^2}{R^2+|z'|^2}(M'+3) \left( (a'+3) - \left( \frac{|z'|^2}{R^2+|z'|^2}(M'+5) + v' \right) \right) \notag \\
    &- \frac{R^2-|z'|^2}{R^2+|z'|^2}\frac{R^2}{R^2+|z'|^2}\frac{|z'|^2}{R^2+|z'|^2} (M'+5)
    \biggr\} \notag \\
    &\times \left( i \frac{1}{R} \right)^2 C^{\prime(M'+6,a'+3)} \left( \frac{R^2}{R^2+|z'|^2} \right)^{\frac{M'-1}{2}} \left( \frac{|z'|}{R} \right)^{-v'} \left( \frac{z'}{R} \right)^{a'},
    \label{eq:+3S2}
\end{align}
\begin{align}
    &{\tilde{\psi}}_{S^2,+,3}^{\prime\,M',a'}(z'=\frac{r}{N+1}e^{i\varphi}, R=\frac{r}{\sqrt{N^2-1}}) \notag \\
    =& \biggl\{ \prod_{k=1}^{3} \left( (a'+k) - \left( \frac{N-1}{2N}(M'+2k-1) + v' \right) \right) \notag \\
    &- \frac{N+1}{2N}\frac{N-1}{2N} (M'+5) \left( (a'+1) - \left( \frac{N-1}{2N} (M'+1) + v' \right) \right) \notag \\
    &- \frac{N+1}{2N}\frac{N-1}{2N} (M'+5) \left( (a'+1) - \left( \frac{N-1}{2N} (M'+3) + v' \right) \right) \notag \\
    &- \frac{N+1}{2N}\frac{N-1}{2N} (M'+3) \left( (a'+1) - \left( \frac{N-1}{2N} (M'+5) + v' \right) \right) \notag \\
    &- \frac{1}{N}\frac{N+1}{2N}\frac{N-1}{2N} (M'+5)
    \biggr\} \notag \\
    &\times \left( i \frac{1}{R} \right)^2 C^{\prime(M'+6,a'+3)} \left( \frac{N+1}{2N} \right)^{\frac{M'-1}{2}} \left( \frac{N-1}{N+1} \right)^{\frac{\ell-v'}{2}} e^{ia'\varphi}.
    \label{eq:+3S2boundary}
\end{align}
In this way, wave functions of level $n$ with the positive and negative chiralities are given in Eqs.~(\ref{eq:psitildeS2+n}) and (\ref{eq:psitildeS2-n}), respectively.


\bibliography{ref}{}

@article{Dixon:1985jw,
    author = "Dixon, Lance J. and Harvey, Jeffrey A. and Vafa, C. and Witten, Edward",
    editor = "Schellekens, B.",
    title = "{Strings on Orbifolds}",
    reportNumber = "PRINT-85-0616 (PRINCETON)",
    doi = "10.1016/0550-3213(85)90593-0",
    journal = "Nucl. Phys. B",
    volume = "261",
    pages = "678--686",
    year = "1985"
}

@article{Dixon:1986jc,
    author = "Dixon, Lance J. and Harvey, Jeffrey A. and Vafa, C. and Witten, Edward",
    title = "{Strings on Orbifolds. 2.}",
    reportNumber = "PRINT-86-0246 (PRINCETON)",
    doi = "10.1016/0550-3213(86)90287-7",
    journal = "Nucl. Phys. B",
    volume = "274",
    pages = "285--314",
    year = "1986"
}

@article{Bachas:1995ik,
    author = "Bachas, C.",
    title = "{A Way to break supersymmetry}",
    eprint = "hep-th/9503030",
    archivePrefix = "arXiv",
    reportNumber = "CPTH-R349-0395, CPTH-RR349-0395, CPTH-RR.349.0395",
    month = "3",
    year = "1995"
}

@article{Blumenhagen:2000wh,
    author = "Blumenhagen, Ralph and Goerlich, Lars and Kors, Boris and Lust, Dieter",
    title = "{Noncommutative compactifications of type I strings on tori with magnetic background flux}",
    eprint = "hep-th/0007024",
    archivePrefix = "arXiv",
    reportNumber = "HUB-EP-00-27",
    doi = "10.1088/1126-6708/2000/10/006",
    journal = "JHEP",
    volume = "10",
    pages = "006",
    year = "2000"
}

@article{Angelantonj:2000hi,
    author = "Angelantonj, C. and Antoniadis, Ignatios and Dudas, E. and Sagnotti, A.",
    title = "{Type I strings on magnetized orbifolds and brane transmutation}",
    eprint = "hep-th/0007090",
    archivePrefix = "arXiv",
    reportNumber = "LPTENS-00-27, CPTH-S074-0600, LPT-ORSAY-00-61, ROM2F-2000-23, CERN-TH-2000-183",
    doi = "10.1016/S0370-2693(00)00907-2",
    journal = "Phys. Lett. B",
    volume = "489",
    pages = "223--232",
    year = "2000"
}

@article{Blumenhagen:2000ea,
    author = "Blumenhagen, Ralph and Kors, Boris and Lust, Dieter",
    title = "{Type I strings with F flux and B flux}",
    eprint = "hep-th/0012156",
    archivePrefix = "arXiv",
    reportNumber = "HU-EP-00-61",
    doi = "10.1088/1126-6708/2001/02/030",
    journal = "JHEP",
    volume = "02",
    pages = "030",
    year = "2001"
}

@article{Cremades:2004wa,
    author = "Cremades, D. and Ibanez, L. E. and Marchesano, F.",
    title = "{Computing Yukawa couplings from magnetized extra dimensions}",
    eprint = "hep-th/0404229",
    archivePrefix = "arXiv",
    reportNumber = "FTUAM-04-7, IFT-UAM-CSIC-04-15, MAD-TH-04-4",
    doi = "10.1088/1126-6708/2004/05/079",
    journal = "JHEP",
    volume = "05",
    pages = "079",
    year = "2004"
}

@article{Abe:2008fi,
    author = "Abe, Hiroyuki and Kobayashi, Tatsuo and Ohki, Hiroshi",
    title = "{Magnetized orbifold models}",
    eprint = "0806.4748",
    archivePrefix = "arXiv",
    primaryClass = "hep-th",
    reportNumber = "YITP-08-48, KUNS-2145",
    doi = "10.1088/1126-6708/2008/09/043",
    journal = "JHEP",
    volume = "09",
    pages = "043",
    year = "2008"
}

@article{Abe:2013bca,
    author = "Abe, Tomo-Hiro and Fujimoto, Yukihiro and Kobayashi, Tatsuo and Miura, Takashi and Nishiwaki, Kenji and Sakamoto, Makoto",
    title = "{$Z_N$ twisted orbifold models with magnetic flux}",
    eprint = "1309.4925",
    archivePrefix = "arXiv",
    primaryClass = "hep-th",
    reportNumber = "KOBE-TH-13-07, KUNS-2463, HRI-P-13-09-001, RECAPP-HRI-2013-019",
    doi = "10.1007/JHEP01(2014)065",
    journal = "JHEP",
    volume = "01",
    pages = "065",
    year = "2014"
}

@article{Abe:2014noa,
    author = "Abe, Tomo-hiro and Fujimoto, Yukihiro and Kobayashi, Tatsuo and Miura, Takashi and Nishiwaki, Kenji and Sakamoto, Makoto",
    title = "{Operator analysis of physical states on magnetized $T^{2}/Z_{N}$ orbifolds}",
    eprint = "1409.5421",
    archivePrefix = "arXiv",
    primaryClass = "hep-th",
    reportNumber = "EHOU-14-003, HRI-P-14-02-001, KOBE-TH-14-02, KUNS-2485, OU-HET-827, RECAPP-HRI-2014-004",
    doi = "10.1016/j.nuclphysb.2014.11.022",
    journal = "Nucl. Phys. B",
    volume = "890",
    pages = "442--480",
    year = "2014"
}

@article{Kobayashi:2017dyu,
    author = "Kobayashi, Tatsuo and Nagamoto, Satoshi",
    title = "{Zero-modes on orbifolds : magnetized orbifold models by modular transformation}",
    eprint = "1709.09784",
    archivePrefix = "arXiv",
    primaryClass = "hep-th",
    reportNumber = "EPHOU-17-013",
    doi = "10.1103/PhysRevD.96.096011",
    journal = "Phys. Rev. D",
    volume = "96",
    number = "9",
    pages = "096011",
    year = "2017"
}

@article{Sakamoto:2020pev,
    author = "Sakamoto, Makoto and Takeuchi, Maki and Tatsuta, Yoshiyuki",
    title = "{Zero-mode counting formula and zeros in orbifold compactifications}",
    eprint = "2004.05570",
    archivePrefix = "arXiv",
    primaryClass = "hep-th",
    reportNumber = "KOBE-TH-20-03, DESY-20-063, DESY 20-063",
    doi = "10.1103/PhysRevD.102.025008",
    journal = "Phys. Rev. D",
    volume = "102",
    number = "2",
    pages = "025008",
    year = "2020"
}

@article{Kikuchi:2022lfv,
    author = "Kikuchi, Shota and Kobayashi, Tatsuo and Nasu, Kaito and Uchida, Hikaru",
    title = "{Classifications of magnetized T$^{4}$ and T$^{4}$/Z$_{2}$ orbifold models}",
    eprint = "2203.01649",
    archivePrefix = "arXiv",
    primaryClass = "hep-th",
    reportNumber = "EPHOU-22-004",
    doi = "10.1007/JHEP08(2022)256",
    journal = "JHEP",
    volume = "08",
    pages = "256",
    year = "2022"
}

@article{Kikuchi:2022psj,
    author = "Kikuchi, Shota and Kobayashi, Tatsuo and Nasu, Kaito and Takada, Shohei and Uchida, Hikaru",
    title = "{Number of zero-modes on magnetized T$^{4}$/Z$_{N}$ orbifolds analyzed by modular transformation}",
    eprint = "2211.07813",
    archivePrefix = "arXiv",
    primaryClass = "hep-th",
    doi = "10.1007/JHEP06(2023)013",
    journal = "JHEP",
    volume = "06",
    pages = "013",
    year = "2023"
}

@article{Kikuchi:2023awm,
    author = "Kikuchi, Shota and Kobayashi, Tatsuo and Nasu, Kaito and Takada, Shohei and Uchida, Hikaru",
    title = "{Zero-modes in magnetized T6/ZN orbifold models through Sp(6,Z) modular symmetry}",
    eprint = "2305.16709",
    archivePrefix = "arXiv",
    primaryClass = "hep-th",
    reportNumber = "EPHOU-23-009",
    doi = "10.1103/PhysRevD.108.036005",
    journal = "Phys. Rev. D",
    volume = "108",
    number = "3",
    pages = "036005",
    year = "2023"
}

@article{Abe:2008sx,
    author = "Abe, Hiroyuki and Choi, Kang-Sin and Kobayashi, Tatsuo and Ohki, Hiroshi",
    title = "{Three generation magnetized orbifold models}",
    eprint = "0812.3534",
    archivePrefix = "arXiv",
    primaryClass = "hep-th",
    reportNumber = "TU-834, KUNS-2173",
    doi = "10.1016/j.nuclphysb.2009.02.002",
    journal = "Nucl. Phys. B",
    volume = "814",
    pages = "265--292",
    year = "2009"
}

@article{Abe:2015yva,
    author = "Abe, Tomo-hiro and Fujimoto, Yukihiro and Kobayashi, Tatsuo and Miura, Takashi and Nishiwaki, Kenji and Sakamoto, Makoto and Tatsuta, Yoshiyuki",
    title = "{Classification of three-generation models on magnetized orbifolds}",
    eprint = "1501.02787",
    archivePrefix = "arXiv",
    primaryClass = "hep-ph",
    reportNumber = "EPHOH-15-001, KIAS-P15002, KOBE-TH-15-01, KUNS-2535, OU-HET-848, WU-HEP-15-01",
    doi = "10.1016/j.nuclphysb.2015.03.004",
    journal = "Nucl. Phys. B",
    volume = "894",
    pages = "374--406",
    year = "2015"
}

@article{Hoshiya:2020hki,
    author = "Hoshiya, Kouki and Kikuchi, Shota and Kobayashi, Tatsuo and Ogawa, Yuya and Uchida, Hikaru",
    title = "{Classification of three-generation models by orbifolding magnetized $T^2 \times T^2$}",
    eprint = "2012.00751",
    archivePrefix = "arXiv",
    primaryClass = "hep-th",
    reportNumber = "EPHOU-20-013",
    doi = "10.1093/ptep/ptab024",
    journal = "PTEP",
    volume = "2021",
    number = "3",
    pages = "033B05",
    year = "2021"
}

@article{Fujimoto:2016zjs,
    author = "Fujimoto, Yukihiro and Kobayashi, Tatsuo and Nishiwaki, Kenji and Sakamoto, Makoto and Tatsuta, Yoshiyuki",
    title = "{Comprehensive analysis of Yukawa hierarchies on $T^2/Z_N$ with magnetic fluxes}",
    eprint = "1605.00140",
    archivePrefix = "arXiv",
    primaryClass = "hep-ph",
    reportNumber = "EPHOH-16-004, KIAS-P16032, KOBE-TH-16-03, WU-HEP-16-07",
    doi = "10.1103/PhysRevD.94.035031",
    journal = "Phys. Rev. D",
    volume = "94",
    number = "3",
    pages = "035031",
    year = "2016"
}

@article{Abe:2009dr,
    author = "Abe, Hiroyuki and Choi, Kang-Sin and Kobayashi, Tatsuo and Ohki, Hiroshi",
    title = "{Higher Order Couplings in Magnetized Brane Models}",
    eprint = "0903.3800",
    archivePrefix = "arXiv",
    primaryClass = "hep-th",
    reportNumber = "KUNS-2196, TU-843",
    doi = "10.1088/1126-6708/2009/06/080",
    journal = "JHEP",
    volume = "06",
    pages = "080",
    year = "2009"
}

@article{Kobayashi:2015siy,
    author = "Kobayashi, Tatsuo and Tatsuta, Yoshiyuki and Uemura, Shohei",
    title = "{Majorana neutrino mass structure induced by rigid instantons on toroidal orbifold}",
    eprint = "1511.09256",
    archivePrefix = "arXiv",
    primaryClass = "hep-ph",
    reportNumber = "KUNS-2596, EPHOU-15-018, WU-HEP-15-22",
    doi = "10.1103/PhysRevD.93.065029",
    journal = "Phys. Rev. D",
    volume = "93",
    number = "6",
    pages = "065029",
    year = "2016"
}

@article{Hoshiya:2021nux,
    author = "Hoshiya, Kouki and Kikuchi, Shota and Kobayashi, Tatsuo and Nasu, Kaito and Uchida, Hikaru and Uemura, Shohei",
    title = "{Majorana neutrino masses by D-brane instanton effects in magnetized orbifold models}",
    eprint = "2103.07147",
    archivePrefix = "arXiv",
    primaryClass = "hep-th",
    reportNumber = "EPHOU-21-005",
    doi = "10.1093/ptep/ptab152",
    journal = "PTEP",
    volume = "2022",
    number = "1",
    pages = "013B04",
    year = "2022"
}

@article{Jeric:2025exr,
    author = "Jeric, Tim and Kobayashi, Tatsuo and Nasu, Kaito and Takada, Shohei",
    title = "{Generation structures and Yukawa couplings in magnetized T2g/ZN models}",
    eprint = "2507.05645",
    archivePrefix = "arXiv",
    primaryClass = "hep-th",
    reportNumber = "EPHOU-25-010",
    doi = "10.1103/rq7n-ypc1",
    journal = "Phys. Rev. D",
    volume = "112",
    number = "7",
    pages = "076033",
    year = "2025"
}

@article{Abe:2012fj,
    author = "Abe, Hiroyuki and Kobayashi, Tatsuo and Ohki, Hiroshi and Oikawa, Akane and Sumita, Keigo",
    title = "{Phenomenological aspects of 10D SYM theory with magnetized extra dimensions}",
    eprint = "1211.4317",
    archivePrefix = "arXiv",
    primaryClass = "hep-ph",
    reportNumber = "WU-HEP-12-06, KUNS-2421",
    doi = "10.1016/j.nuclphysb.2013.01.014",
    journal = "Nucl. Phys. B",
    volume = "870",
    pages = "30--54",
    year = "2013"
}

@article{Abe:2014vza,
    author = "Abe, Hiroyuki and Kobayashi, Tatsuo and Sumita, Keigo and Tatsuta, Yoshiyuki",
    title = "{Gaussian Froggatt-Nielsen mechanism on magnetized orbifolds}",
    eprint = "1405.5012",
    archivePrefix = "arXiv",
    primaryClass = "hep-ph",
    reportNumber = "WU-HEP-14-06, EPHOU-14-012",
    doi = "10.1103/PhysRevD.90.105006",
    journal = "Phys. Rev. D",
    volume = "90",
    number = "10",
    pages = "105006",
    year = "2014"
}

@article{Kobayashi:2016qag,
    author = "Kobayashi, Tatsuo and Nishiwaki, Kenji and Tatsuta, Yoshiyuki",
    title = "{CP-violating phase on magnetized toroidal orbifolds}",
    eprint = "1609.08608",
    archivePrefix = "arXiv",
    primaryClass = "hep-th",
    reportNumber = "EPHOU-16-018, KIAS-P16072, WU-HEP-16-19",
    doi = "10.1007/JHEP04(2017)080",
    journal = "JHEP",
    volume = "04",
    pages = "080",
    year = "2017"
}

@article{Buchmuller:2017vho,
    author = "Buchmuller, Wilfried and Schweizer, Julian",
    title = "{Flavor mixings in flux compactifications}",
    eprint = "1701.06935",
    archivePrefix = "arXiv",
    primaryClass = "hep-ph",
    reportNumber = "DESY-16-238",
    doi = "10.1103/PhysRevD.95.075024",
    journal = "Phys. Rev. D",
    volume = "95",
    number = "7",
    pages = "075024",
    year = "2017"
}

@article{Buchmuller:2017vut,
    author = "Buchmuller, Wilfried and Patel, Ketan M.",
    title = "{Flavor physics without flavor symmetries}",
    eprint = "1712.06862",
    archivePrefix = "arXiv",
    primaryClass = "hep-ph",
    reportNumber = "DESY-17-220",
    doi = "10.1103/PhysRevD.97.075019",
    journal = "Phys. Rev. D",
    volume = "97",
    number = "7",
    pages = "075019",
    year = "2018"
}

@article{Kikuchi:2021yog,
    author = "Kikuchi, Shota and Kobayashi, Tatsuo and Ogawa, Yuya and Uchida, Hikaru",
    title = "{Yukawa textures in modular symmetric vacuum of magnetized orbifold models}",
    eprint = "2112.01680",
    archivePrefix = "arXiv",
    primaryClass = "hep-ph",
    reportNumber = "EPHOU-21-020",
    doi = "10.1093/ptep/ptac035",
    journal = "PTEP",
    volume = "2022",
    number = "3",
    pages = "033B10",
    year = "2022"
}

@article{Kikuchi:2022geu,
    author = "Kikuchi, Shota and Kobayashi, Tatsuo and Tanimoto, Morimitsu and Uchida, Hikaru",
    title = "{Mass matrices with CP phase in modular flavor symmetry}",
    eprint = "2206.08538",
    archivePrefix = "arXiv",
    primaryClass = "hep-ph",
    reportNumber = "EPHOU-22-009",
    doi = "10.1093/ptep/ptac141",
    journal = "PTEP",
    volume = "2022",
    number = "11",
    pages = "113B07",
    year = "2022"
}

@article{Hoshiya:2022qvr,
    author = "Hoshiya, Kouki and Kikuchi, Shota and Kobayashi, Tatsuo and Uchida, Hikaru",
    title = "{Quark and lepton flavor structure in magnetized orbifold models at residual modular symmetric points}",
    eprint = "2209.07249",
    archivePrefix = "arXiv",
    primaryClass = "hep-ph",
    reportNumber = "EPHOU-22-018",
    doi = "10.1103/PhysRevD.106.115003",
    journal = "Phys. Rev. D",
    volume = "106",
    number = "11",
    pages = "115003",
    year = "2022"
}

@article{Kobayashi:2018rad,
    author = "Kobayashi, Tatsuo and Nagamoto, Satoshi and Takada, Shintaro and Tamba, Shio and Tatsuishi, Takuya H.",
    title = "{Modular symmetry and non-Abelian discrete flavor symmetries in string compactification}",
    eprint = "1804.06644",
    archivePrefix = "arXiv",
    primaryClass = "hep-th",
    reportNumber = "EPHOU-18-003",
    doi = "10.1103/PhysRevD.97.116002",
    journal = "Phys. Rev. D",
    volume = "97",
    number = "11",
    pages = "116002",
    year = "2018"
}

@article{Kikuchi:2020frp,
    author = "Kikuchi, Shota and Kobayashi, Tatsuo and Takada, Shintaro and Tatsuishi, Takuya H. and Uchida, Hikaru",
    title = "{Revisiting modular symmetry in magnetized torus and orbifold compactifications}",
    eprint = "2005.12642",
    archivePrefix = "arXiv",
    primaryClass = "hep-th",
    reportNumber = "EPHOU-20-005",
    doi = "10.1103/PhysRevD.102.105010",
    journal = "Phys. Rev. D",
    volume = "102",
    number = "10",
    pages = "105010",
    year = "2020"
}

@article{Kikuchi:2021ogn,
    author = "Kikuchi, Shota and Kobayashi, Tatsuo and Uchida, Hikaru",
    title = "{Modular flavor symmetries of three-generation modes on magnetized toroidal orbifolds}",
    eprint = "2101.00826",
    archivePrefix = "arXiv",
    primaryClass = "hep-th",
    reportNumber = "EPHOU-21-001",
    doi = "10.1103/PhysRevD.104.065008",
    journal = "Phys. Rev. D",
    volume = "104",
    number = "6",
    pages = "065008",
    year = "2021"
}

@article{Kobayashi:2018bff,
    author = "Kobayashi, Tatsuo and Tamba, Shio",
    title = "{Modular forms of finite modular subgroups from magnetized D-brane models}",
    eprint = "1811.11384",
    archivePrefix = "arXiv",
    primaryClass = "hep-th",
    reportNumber = "EPHOU-18-014",
    doi = "10.1103/PhysRevD.99.046001",
    journal = "Phys. Rev. D",
    volume = "99",
    number = "4",
    pages = "046001",
    year = "2019"
}

@article{Kikuchi:2022bkn,
    author = "Kikuchi, Shota and Kobayashi, Tatsuo and Nasu, Kaito and Uchida, Hikaru and Uemura, Shohei",
    title = "{Modular symmetry anomaly and nonperturbative neutrino mass terms in magnetized orbifold models}",
    eprint = "2202.05425",
    archivePrefix = "arXiv",
    primaryClass = "hep-th",
    reportNumber = "EPHOU-22-003",
    doi = "10.1103/PhysRevD.105.116002",
    journal = "Phys. Rev. D",
    volume = "105",
    number = "11",
    pages = "116002",
    year = "2022"
}

@article{Kikuchi:2023clx,
    author = "Kikuchi, Shota and Kobayashi, Tatsuo and Nasu, Kaito and Otsuka, Hajime and Takada, Shohei and Uchida, Hikaru",
    title = "{Remark on modular weights in low-energy effective field theory from type II string theory}",
    eprint = "2301.10356",
    archivePrefix = "arXiv",
    primaryClass = "hep-th",
    reportNumber = "EPHOU-23-003, KYUSHU-HET-255",
    doi = "10.1007/JHEP04(2023)003",
    journal = "JHEP",
    volume = "04",
    pages = "003",
    year = "2023"
}

@article{Kikuchi:2023awe,
    author = "Kikuchi, Shota and Kobayashi, Tatsuo and Nasu, Kaito and Takada, Shohei and Uchida, Hikaru",
    title = "{Modular symmetry in magnetized T2g torus and orbifold models}",
    eprint = "2309.16447",
    archivePrefix = "arXiv",
    primaryClass = "hep-th",
    reportNumber = "EPHOU-23-016",
    doi = "10.1103/PhysRevD.109.065011",
    journal = "Phys. Rev. D",
    volume = "109",
    number = "6",
    pages = "065011",
    year = "2024"
}

@article{Kobayashi:2024ysa,
    author = "Kobayashi, Tatsuo and Nasu, Kaito and Nishida, Ryusei and Otsuka, Hajime and Takada, Shohei",
    title = "{Flavor symmetries from modular subgroups in magnetized compactifications}",
    eprint = "2409.02458",
    archivePrefix = "arXiv",
    primaryClass = "hep-th",
    reportNumber = "EPHOU-24-012, KYUSHU-HET-293",
    doi = "10.1007/JHEP12(2024)128",
    journal = "JHEP",
    volume = "12",
    pages = "128",
    year = "2024"
}

@article{Kobayashi:2024hkk,
    author = "Kobayashi, Tatsuo and Otsuka, Hajime and Takada, Shohei and Uchida, Hikaru",
    title = "{Modular symmetry of localized modes}",
    eprint = "2410.05788",
    archivePrefix = "arXiv",
    primaryClass = "hep-th",
    reportNumber = "EPHOU-24-015, KYUSHU-HET-299",
    doi = "10.1103/PhysRevD.110.125013",
    journal = "Phys. Rev. D",
    volume = "110",
    number = "12",
    pages = "125013",
    year = "2024"
}

@article{Jeric:2025iwk,
    author = "Jeric, Tim and Kobayashi, Tatsuo and Otsuka, Hajime and Takeuchi, Maki and Uchida, Hikaru",
    title = "{Modular weights of wave functions on magnetized torus}",
    eprint = "2512.18574",
    archivePrefix = "arXiv",
    primaryClass = "hep-th",
    reportNumber = "EPHOU-25-020, KYUSHU-HET-346",
    month = "12",
    year = "2025"
}

@article{Kobayashi:2024yqq,
    author = "Kobayashi, Tatsuo and Otsuka, Hajime",
    title = "{Non-invertible flavor symmetries in magnetized extra dimensions}",
    eprint = "2408.13984",
    archivePrefix = "arXiv",
    primaryClass = "hep-th",
    reportNumber = "EPHOU-24-010, KYUSHU-HET-292",
    doi = "10.1007/JHEP11(2024)120",
    journal = "JHEP",
    volume = "11",
    pages = "120",
    year = "2024"
}

@article{Hamada:2012wj,
    author = "Hamada, Yuta and Kobayashi, Tatsuo",
    title = "{Massive Modes in Magnetized Brane Models}",
    eprint = "1207.6867",
    archivePrefix = "arXiv",
    primaryClass = "hep-th",
    reportNumber = "KUNS-2409",
    doi = "10.1143/PTP.128.903",
    journal = "Prog. Theor. Phys.",
    volume = "128",
    pages = "903--923",
    year = "2012"
}

@article{GrootNibbelink:2007lua,
    author = "Groot Nibbelink, S. and Trapletti, M. and Walter, M.",
    title = "{Resolutions of C**n/Z(n) Orbifolds, their U(1) Bundles, and Applications to String Model Building}",
    eprint = "hep-th/0701227",
    archivePrefix = "arXiv",
    reportNumber = "HD-THEP-07-03, SIAS-CMTP-07-1",
    doi = "10.1088/1126-6708/2007/03/035",
    journal = "JHEP",
    volume = "03",
    pages = "035",
    year = "2007"
}

@article{Leung:2019oln,
    author = "Leung, Pompey and Otsuka, Hajime",
    title = "{Heterotic Stringy Corrections to Metrics of Toroidal Orbifolds and Their Resolutions}",
    eprint = "1903.12144",
    archivePrefix = "arXiv",
    primaryClass = "hep-th",
    reportNumber = "WU-HEP-19-04",
    doi = "10.1103/PhysRevD.99.126011",
    journal = "Phys. Rev. D",
    volume = "99",
    number = "12",
    pages = "126011",
    year = "2019"
}

@article{Eguchi:1978xp,
    author = "Eguchi, Tohru and Hanson, Andrew J.",
    title = "{Asymptotically Flat Selfdual Solutions to Euclidean Gravity}",
    reportNumber = "SLAC-PUB-2087, LBL-7273",
    doi = "10.1016/0370-2693(78)90566-X",
    journal = "Phys. Lett. B",
    volume = "74",
    pages = "249--251",
    year = "1978"
}

@article{Kobayashi:2019fma,
    author = "Kobayashi, Tatsuo and Otsuka, Hajime and Uchida, Hikaru",
    title = "{Wavefunctions and Yukawa couplings on resolutions of T$^{2}$/{\ensuremath{\mathbb{Z}}}$_{N}$ orbifolds}",
    eprint = "1904.02867",
    archivePrefix = "arXiv",
    primaryClass = "hep-th",
    reportNumber = "EPHOU-19-006, KEK-TH-2118",
    doi = "10.1007/JHEP08(2019)046",
    journal = "JHEP",
    volume = "08",
    pages = "046",
    year = "2019"
}

@article{Kobayashi:2019gyl,
    author = "Kobayashi, Tatsuo and Otsuka, Hajime and Uchida, Hikaru",
    title = "{Flavor structure of magnetized $T^2/\mathbb{Z}_2$ blow-up models}",
    eprint = "1911.01930",
    archivePrefix = "arXiv",
    primaryClass = "hep-ph",
    reportNumber = "EPHOU-19-0014, KEK-TH-2166",
    doi = "10.1007/JHEP03(2020)042",
    journal = "JHEP",
    volume = "03",
    pages = "042",
    year = "2020"
}

@article{Kobayashi:2022tti,
    author = "Kobayashi, Tatsuo and Otsuka, Hajime and Sakamoto, Makoto and Takeuchi, Maki and Tatsuta, Yoshiyuki and Uchida, Hikaru",
    title = "{Index theorem on magnetized blow-up manifold of T2/ZN}",
    eprint = "2211.04595",
    archivePrefix = "arXiv",
    primaryClass = "hep-th",
    reportNumber = "EPHOU-22-019, KYUSHU-HET-249, KOBE-TH-22-05",
    doi = "10.1103/PhysRevD.107.075032",
    journal = "Phys. Rev. D",
    volume = "107",
    number = "7",
    pages = "075032",
    year = "2023"
}

@article{Conlon:2008qi,
    author = "Conlon, Joseph P. and Maharana, Anshuman and Quevedo, Fernando",
    title = "{Wave Functions and Yukawa Couplings in Local String Compactifications}",
    eprint = "0807.0789",
    archivePrefix = "arXiv",
    primaryClass = "hep-th",
    reportNumber = "DAMTP-2008-33",
    doi = "10.1088/1126-6708/2008/09/104",
    journal = "JHEP",
    volume = "09",
    pages = "104",
    year = "2008"
}

@article{Dolan:2020sjq,
    author = "Dolan, Brian P. and Hunter-McCabe, Aonghus",
    title = "{Ground state wave functions for the quantum Hall effect on a sphere and the Atiyah-Singer index theorem}",
    eprint = "2001.02208",
    archivePrefix = "arXiv",
    primaryClass = "hep-th",
    reportNumber = "DIAS-STP-20-01",
    doi = "10.1088/1751-8121/ab85e1",
    journal = "J. Phys. A",
    volume = "53",
    number = "21",
    pages = "215306",
    year = "2020"
}

@article{Lee:2003mc,
    author = "Lee, Hyun Min and Nilles, Hans Peter and Zucker, Max",
    title = "{Spontaneous localization of bulk fields: The Six-dimensional case}",
    eprint = "hep-th/0309195",
    archivePrefix = "arXiv",
    doi = "10.1016/j.nuclphysb.2003.12.031",
    journal = "Nucl. Phys. B",
    volume = "680",
    pages = "177--198",
    year = "2004"
}

@article{Buchmuller:2015eya,
    author = "Buchmuller, Wilfried and Dierigl, Markus and Ruehle, Fabian and Schweizer, Julian",
    title = "{Chiral fermions and anomaly cancellation on orbifolds with Wilson lines and flux}",
    eprint = "1506.05771",
    archivePrefix = "arXiv",
    primaryClass = "hep-th",
    reportNumber = "DESY-15-068",
    doi = "10.1103/PhysRevD.92.105031",
    journal = "Phys. Rev. D",
    volume = "92",
    number = "10",
    pages = "105031",
    year = "2015"
}

@article{Buchmuller:2018lkz,
    author = "Buchmuller, Wilfried and Dierigl, Markus and Tatsuta, Yoshiyuki",
    title = "{Magnetized orbifolds and localized flux}",
    eprint = "1810.06362",
    archivePrefix = "arXiv",
    primaryClass = "hep-th",
    reportNumber = "DESY 18-167, DESY-18-167",
    doi = "10.1016/j.aop.2018.12.006",
    journal = "Annals Phys.",
    volume = "401",
    pages = "91--115",
    year = "2019"
}

@book{Polchinski:1998rq,
    author = "Polchinski, J.",
    title = "{String theory. Vol. 1: An introduction to the bosonic string}",
    doi = "10.1017/CBO9780511816079",
    isbn = "978-0-511-25227-3, 978-0-521-67227-6, 978-0-521-63303-1",
    publisher = "Cambridge University Press",
    series = "Cambridge Monographs on Mathematical Physics",
    month = "12",
    year = "2007"
}

@article{Ohki:2020bpo,
    author = "Ohki, Hiroshi and Uemura, Shohei and Watanabe, Risa",
    title = "{Modular flavor symmetry on a magnetized torus}",
    eprint = "2003.04174",
    archivePrefix = "arXiv",
    primaryClass = "hep-th",
    doi = "10.1103/PhysRevD.102.085008",
    journal = "Phys. Rev. D",
    volume = "102",
    number = "8",
    pages = "085008",
    year = "2020"
}

@article{Berasaluce-Gonzalez:2012abm,
    author = "Berasaluce-Gonzalez, M. and Camara, P. G. and Marchesano, F. and Regalado, D. and Uranga, A. M.",
    title = "{Non-Abelian discrete gauge symmetries in 4d string models}",
    eprint = "1206.2383",
    archivePrefix = "arXiv",
    primaryClass = "hep-th",
    reportNumber = "IFT-UAM-CSIC-12-53",
    doi = "10.1007/JHEP09(2012)059",
    journal = "JHEP",
    volume = "09",
    pages = "059",
    year = "2012"
}

@article{Ding:2020zxw,
    author = "Ding, Gui-Jun and Feruglio, Ferruccio and Liu, Xiang-Gan",
    title = "{Automorphic Forms and Fermion Masses}",
    eprint = "2010.07952",
    archivePrefix = "arXiv",
    primaryClass = "hep-th",
    reportNumber = "USTC-ICTS/PCFT-20-28",
    doi = "10.1007/JHEP01(2021)037",
    journal = "JHEP",
    volume = "01",
    pages = "037",
    year = "2021"
}

@article{Goddard:1986bp,
    author = "Goddard, Peter and Olive, David I.",
    title = "{Kac-Moody and Virasoro Algebras in Relation to Quantum Physics}",
    reportNumber = "DAMTP-86-5",
    doi = "10.1142/S0217751X86000149",
    journal = "Int. J. Mod. Phys. A",
    volume = "1",
    pages = "303",
    year = "1986"
}

@article{Candelas:1985en,
    author = "Candelas, P. and Horowitz, Gary T. and Strominger, Andrew and Witten, Edward",
    title = "{Vacuum configurations for superstrings}",
    reportNumber = "NSF-ITP-84-170",
    doi = "10.1016/0550-3213(85)90602-9",
    journal = "Nucl. Phys. B",
    volume = "258",
    pages = "46--74",
    year = "1985"
}

@inbook{Feruglio:2017spp,
    author = "Feruglio, Ferruccio",
    editor = "Levy, Aharon and Forte, Stefano and Ridolfi, Giovanni",
    title = "{Are neutrino masses modular forms?}",
    booktitle = "{From My Vast Repertoire ...}: {Guido Altarelli's Legacy}",
    eprint = "1706.08749",
    archivePrefix = "arXiv",
    primaryClass = "hep-ph",
    reportNumber = "DFPD-2017-TH-09",
    doi = "10.1142/9789813238053_0012",
    pages = "227--266",
    year = "2019"
}

@article{Kobayashi:2018vbk,
    author = "Kobayashi, Tatsuo and Tanaka, Kentaro and Tatsuishi, Takuya H.",
    title = "{Neutrino mixing from finite modular groups}",
    eprint = "1803.10391",
    archivePrefix = "arXiv",
    primaryClass = "hep-ph",
    reportNumber = "HPHOU-18-002",
    doi = "10.1103/PhysRevD.98.016004",
    journal = "Phys. Rev. D",
    volume = "98",
    number = "1",
    pages = "016004",
    year = "2018"
}

@article{Penedo:2018nmg,
    author = "Penedo, J. T. and Petcov, S. T.",
    title = "{Lepton Masses and Mixing from Modular $S_4$ Symmetry}",
    eprint = "1806.11040",
    archivePrefix = "arXiv",
    primaryClass = "hep-ph",
    reportNumber = "SISSA 25/2018/FISI, IPMU18-0121, SISSA-25-2018-FISI",
    doi = "10.1016/j.nuclphysb.2018.12.016",
    journal = "Nucl. Phys. B",
    volume = "939",
    pages = "292--307",
    year = "2019"
}

@article{Kobayashi:2018scp,
    author = "Kobayashi, Tatsuo and Omoto, Naoya and Shimizu, Yusuke and Takagi, Kenta and Tanimoto, Morimitsu and Tatsuishi, Takuya H.",
    title = "{Modular A$_{4}$ invariance and neutrino mixing}",
    eprint = "1808.03012",
    archivePrefix = "arXiv",
    primaryClass = "hep-ph",
    reportNumber = "EPHOU-18-009, HUPD1807",
    doi = "10.1007/JHEP11(2018)196",
    journal = "JHEP",
    volume = "11",
    pages = "196",
    year = "2018"
}

@article{Novichkov:2018nkm,
    author = "Novichkov, P. P. and Penedo, J. T. and Petcov, S. T. and Titov, A. V.",
    title = "{Modular A$_{5}$ symmetry for flavour model building}",
    eprint = "1812.02158",
    archivePrefix = "arXiv",
    primaryClass = "hep-ph",
    reportNumber = "SISSA 54/2018/FISI, IPMU18-0202, IPPP/18/105",
    doi = "10.1007/JHEP04(2019)174",
    journal = "JHEP",
    volume = "04",
    pages = "174",
    year = "2019"
}

@article{Kobayashi:2023zzc,
    author = "Kobayashi, Tatsuo and Tanimoto, Morimitsu",
    editor = {Buchalla, Gerhard and L{\"u}st, Dieter and Xing, Zhi-zhong},
    title = "{Modular flavor symmetric models}",
    eprint = "2307.03384",
    archivePrefix = "arXiv",
    primaryClass = "hep-ph",
    doi = "10.1142/S0217751X24410124",
    journal = "Int. J. Mod. Phys. A",
    volume = "39",
    number = "09n10",
    pages = "2441012",
    year = "2024"
}

@article{Ding:2023htn,
    author = "Ding, Gui-Jun and King, Stephen F.",
    title = "{Neutrino mass and mixing with modular symmetry}",
    eprint = "2311.09282",
    archivePrefix = "arXiv",
    primaryClass = "hep-ph",
    doi = "10.1088/1361-6633/ad52a3",
    journal = "Rept. Prog. Phys.",
    volume = "87",
    number = "8",
    pages = "084201",
    year = "2024"
}

@article{Dolan:2003bj,
    author = "Dolan, Brian P.",
    title = "{The Spectrum of the Dirac operator on coset spaces with homogeneous gauge fields}",
    eprint = "hep-th/0304037",
    archivePrefix = "arXiv",
    reportNumber = "DIAS-STP-03-02",
    doi = "10.1088/1126-6708/2003/05/018",
    journal = "JHEP",
    volume = "05",
    pages = "018",
    year = "2003"
}

@article{Dixon:1990pc,
    author = "Dixon, Lance J. and Kaplunovsky, Vadim and Louis, Jan",
    title = "{Moduli dependence of string loop corrections to gauge coupling constants}",
    reportNumber = "SLAC-PUB-5138, UTTG-36-89",
    doi = "10.1016/0550-3213(91)90490-O",
    journal = "Nucl. Phys. B",
    volume = "355",
    pages = "649--688",
    year = "1991"
}

@article{Antoniadis:1992pm,
    author = "Antoniadis, Ignatios and Gava, E. and Narain, K. S. and Taylor, T. R.",
    title = "{Superstring threshold corrections to Yukawa couplings}",
    eprint = "hep-th/9212045",
    archivePrefix = "arXiv",
    reportNumber = "NUB-3057, IC-92-416, CPTH-A185-0892",
    doi = "10.1016/0550-3213(93)90095-7",
    journal = "Nucl. Phys. B",
    volume = "407",
    pages = "706--724",
    year = "1993"
}

@article{Choi:2020dws,
    author = "Choi, Kang-Sin and Kim, Jihn E.",
    title = "{Quarks and Leptons From Orbifolded Superstring}",
    doi = "10.1007/978-3-030-54005-0",
    journal = "Lect. Notes Phys.",
    volume = "954",
    pages = "pp.",
    year = "2020"
}

@article{Aldazabal:2000cn,
    author = "Aldazabal, G. and Franco, S. and Ibanez, Luis E. and Rabadan, R. and Uranga, A. M.",
    title = "{Intersecting brane worlds}",
    eprint = "hep-ph/0011132",
    archivePrefix = "arXiv",
    reportNumber = "CAB-IB-2919500, CERN-TH-2000-320, MIT-CTP-3042, FTUAM-00-23, IFT-UAM-CSIC-00-37",
    doi = "10.1088/1126-6708/2001/02/047",
    journal = "JHEP",
    volume = "02",
    pages = "047",
    year = "2001"
}

@article{Cvetic:2003ch,
    author = "Cvetic, Mirjam and Papadimitriou, Ioannis",
    title = "{Conformal field theory couplings for intersecting D-branes on orientifolds}",
    eprint = "hep-th/0303083",
    archivePrefix = "arXiv",
    reportNumber = "UPR-1028-T, PUPT-2076",
    doi = "10.1103/PhysRevD.70.029903",
    journal = "Phys. Rev. D",
    volume = "68",
    pages = "046001",
    year = "2003",
    note = "[Erratum: Phys.Rev.D 70, 029903 (2004)]"
}

@article{Abel:2003vv,
    author = "Abel, S. A. and Owen, A. W.",
    title = "{Interactions in intersecting brane models}",
    eprint = "hep-th/0303124",
    archivePrefix = "arXiv",
    reportNumber = "IPPP-03-13, DCPT-03-26",
    doi = "10.1016/S0550-3213(03)00370-5",
    journal = "Nucl. Phys. B",
    volume = "663",
    pages = "197--214",
    year = "2003"
}

@article{Hamidi:1986vh,
    author = "Hamidi, Shahram and Vafa, Cumrun",
    title = "{Interactions on Orbifolds}",
    reportNumber = "HUTP-86-A041, CALT-68-1349",
    doi = "10.1016/0550-3213(87)90006-X",
    journal = "Nucl. Phys. B",
    volume = "279",
    pages = "465--513",
    year = "1987"
}

@article{Dixon:1986qv,
    author = "Dixon, Lance J. and Friedan, Daniel and Martinec, Emil J. and Shenker, Stephen H.",
    title = "{The Conformal Field Theory of Orbifolds}",
    reportNumber = "EFI-86-42-CHICAGO",
    doi = "10.1016/0550-3213(87)90676-6",
    journal = "Nucl. Phys. B",
    volume = "282",
    pages = "13--73",
    year = "1987"
}
\bibliographystyle{JHEP} 

\end{document}